\documentclass[fleqn,usenatbib]{mnras}

\usepackage{newtxtext,newtxmath}

\usepackage[T1]{fontenc}

\DeclareRobustCommand{\VAN}[3]{#2}
\let\VANthebibliography\thebibliography
\def\thebibliography{\DeclareRobustCommand{\VAN}[3]{##3}\VANthebibliography}

\newcommand{\gaia}{{\em Gaia}}
\newcommand{\hst}{{\em HST}}
\newcommand{\jwst}{{\em JWST}}
\newcommand{\ngrst}{{\em NGRST}}

\usepackage{graphicx}	
\usepackage{amsmath}	

\title[Ejected supernova companions]{Searching for ejected supernova companions in the era of precise proper motion and radial velocity measurements}

\author[A. A. Chrimes et al.]{A. A. Chrimes,$^{1}$\thanks{E-mail: a.chrimes@astro.ru.nl}
A. J. Levan$^{1,2}$,
J. J. Eldridge$^{3}$,
M. Fraser$^{4}$,
N. Gaspari$^{1}$,
P. J. Groot$^{1,5,6}$,
J. D. Lyman$^{2}$,\newauthor
G. Nelemans$^{1,7,8}$,
E. R. Stanway$^{2}$
and K. Wiersema$^{9}$\\
$^{1}$Department of Astrophysics/IMAPP, Radboud University, PO Box 9010, 6500 GL Nijmegen, The Netherlands \\
$^{2}$Department of Physics, University of Warwick, Gibbet Hill Road, Coventry, UK \\
$^{3}$Department of Physics, University of Auckland, Private Bag 92019, Auckland, New Zealand \\
$^{4}$School of Physics, O’Brien Centre for Science North, University College Dublin, Belfield, Dublin 4, Ireland\\
$^{5}$Inter-University Institute for Data Intensive Astronomy, Department of Astronomy, University of Cape Town, Private Bag X3, Rondebosch 7701, South Africa \\
$^{6}$South African Astronomical Observatory, P.O. Box 9, 7935 Observatory, South Africa \\
$^{7}$Institute for Astronomy, KU Leuven, Leuven, Belgium \\
$^{8}$SRON, Netherlands Institute for Space Research, Sorbonnelaan 2, NL-3584 CA Utrecht, The Netherlands \\
$^{9}$Physics Department, Lancaster University, Lancaster, LA1 4YB, UK \\
}

\date{Accepted XXX. Received YYY; in original form ZZZ}

\pubyear{2023}

\begin{document}
\label{firstpage}
\pagerange{\pageref{firstpage}--\pageref{lastpage}}
\maketitle
\begin{abstract}
The majority of massive stars are born in binaries, and most unbind upon the first supernova. With precise proper motion surveys such as \gaia, it is possible to trace back the motion of stars in the vicinity of young remnants to search for ejected companions. Establishing the fraction of remnants with an ejected companion, and the photometric and kinematic properties of these stars, offers unique insight into supernova progenitor systems. In this paper, we employ binary population synthesis to produce kinematic and photometric predictions for ejected secondary stars. We demonstrate that the unbound neutron star velocity distribution from supernovae in binaries closely traces the input kicks. Therefore, the observed distribution of neutron star velocities should be representative of their natal kicks. We evaluate the probability for any given filter, magnitude limit, minimum measurable proper motion (as a function of magnitude), temporal baseline, distance and extinction that an unbound companion can be associated with a remnant. We compare our predictions with results from previous companion searches, and demonstrate that the current sample of stars ejected by the supernova of their companion can be increased by a factor of $\sim$5--10 with \gaia\ data release 3. Further progress in this area is achievable by leveraging the absolute astrometric precision of \gaia, and by obtaining multiple epochs of deep, high resolution near-infrared imaging with the {\em Hubble Space Telescope}, \jwst\ and next-generation wide-field near-infrared observatories such as {\em Euclid} or the {\it Nancy Grace Roman Space Telescope}.
\end{abstract}

\begin{keywords}
supernovae: general -- stars: kinematics and dynamics -- stars: neutron stars -- proper motions -- techniques: radial velocities
\end{keywords}



\section{Introduction}
Most massive stars are born in binaries or higher order systems. The binary fraction is a strong function of mass, rising from $\sim$0.4 at a Solar mass to $\sim$1 above $10$\,M$_{\odot}$ \citep{2012Sci...337..444S,2014ApJS..215...15S,2017ApJS..230...15M,2020MNRAS.495.4605S}. At the highest masses, triple, quadruple and higher order systems are also common \citep[][and reference therein]{2022arXiv220310066O}. Based on the observed distribution of stellar and binary parameters, such as the initial mass function, mass ratio distribution and orbital periods, and binary interaction modelling, up to a third of massive star binaries - in which at least one component undergoes core-collapse - are expected to merge before the first supernova\footnote{Throughout, we refer exclusively to core-collapse supernovae} \citep{2012Sci...337..444S,2019A&A...624A..66R}. Of the remainder, a large majority are expected to unbind upon the first supernova \citep[e.g. 86$^{+11}_{-22}$ per cent,][]{2019A&A...624A..66R}. Throughout, we only consider single stars and binaries (not triples or higher-order systems).

The velocity distribution of field OB stars has a tail which extends to high velocities in excess of 30\,km\,s$^{-1}$; these stars are referred to as `runaways' \citep{1961BAN....15..265B,2011MNRAS.410..190T,2018A&A...616A.149M}. Stars ejected by the supernova of a companion, but moving slower than 30\,km\,s$^{-1}$, have been termed `walkaways' \citep{2012ASPC..465...65D}. In addition to ejection by companion supernovae, dynamical interactions in clusters and dense star forming regions can also eject stars \citep{1967BOTT....4...86P,2018A&A...619A..78L,2019A&A...625L...2K,2019A&A...624A..66R}.

Uncertainties in velocity distribution and ejection mechanism arise from the local reference frame being hard to define, which is crucial to determine the peculiar velocity in the local standard of rest, and the fraction of dynamically ejected stars possibly constituting a larger fraction of the OB star population than expected \citep{2019A&A...624A..66R}. There also exists a class of `hypervelocity' stars travelling at up to $\sim$1000\,km\,s$^{-1}$, some of which are likely ejected by dynamical interactions in the Galactic centre \citep[e.g.][]{2017MNRAS.469.2151B,2018ApJ...867L...8O,2020MNRAS.497.5344E,2022MNRAS.515..767M}. In this paper, we focus exclusively on walkaways and runaways ejected by supernovae (SNe) in binaries. 

Most massive star binaries interact at some stage of their evolution \citep{1976ApJS...30..273A,1992ApJ...391..246P,2017PASA...34...58E,2019A&A...624A..66R}, for example through mass transfer or tidal interactions. This can influence the properties of both the primary and secondary star. For example, stripping by a companion is thought to be, at least in part, responsible for removing the envelopes of massive stars which explode as type Ibc supernovae \citep{1995PhR...256..173N,1997ARA&A..35..309F,2019NatAs...3..434F,2022arXiv220905283S}. Meanwhile, the secondary accretes mass and can be spun up \citep[e.g.][]{1991A&A...241..419P}, in extreme cases this could lead to unusual evolution \citep[such as homogeneous or quasi-homogeneous evolution, e.g.][]{2011MNRAS.414.3501E,2016MNRAS.458.2634M,2022arXiv220803999G}, or they can be polluted by winds from a stripped-envelope companion \citep{2014A&A...565A..90C}. It has also been suggested that secondaries might appear inflated and red for a short time ($\sim$1000\,yr) after the supernova of a companion, as a result of the impact of the primary ejecta \citep{2015A&A...584A..11L,2018ApJ...864..119H,2021MNRAS.505.2485O}. For these reasons, identifying secondary stars ejected by the supernovae of primary stars promises to reveal much about the binary progenitors of supernovae and their interactions. By tracing the proper motions of stars backwards in time, and searching for paths which intersect supernova remnants, it is possible to search for unbound companions \citep[e.g.][]{2015MNRAS.448.3196D,2017A&A...606A..14B,2019ApJ...871...92F,2020MNRAS.498..899N,2021AN....342..553L}. Such searches are now being carried out, with the groundbreaking capability of \gaia\ \citep[most recently data release 3,][]{2022arXiv220800211G} to measure proper motions at the millarcsecond level for the vast majority of stars down to $\sim$21$^{\rm st}$ magnitude. A combination of dynamical and supernova ejection may result in a small fraction of massive stars which cannot be traced back to a young stellar cluster or association \citep{2010MNRAS.404.1564P,2012MNRAS.424.3037G}, but associating runaways with a supernova remnant may still be possible in these cases (even if it is located outside the progenitor's birth cluster).

Establishing the observed kinematic properties of ejected secondaries may also offer new insight into neutron star natal kicks. Thus far, these constraints have arisen from measurement of the velocity distribution of isolated neutron stars \citep{2005MNRAS.360..974H,2017A&A...608A..57V,2020MNRAS.494.3663I}, or from the systemic velocities, orbital periods and eccentricities of bound systems, either after one supernova \citep[living-degenerate binaries,][]{2019MNRAS.486.4098I,2022arXiv220603904F}, or two \citep[double degenerate binaries,][]{2017ApJ...846..170T}. The velocity distribution of stars ejected by supernovae in binaries is dominated by the pre-supernova orbital velocity of the system \citep{1998A&A...330.1047T}, so measuring this distribution gives insight into the pre-supernova binary parameters.

Going forwards, there are great opportunities to make advances in these areas. With the absolute reference provided by \gaia, we can accurately compare proper motions measured by multiple observatories. This facilitates searches for runaways not only from supernova remnants, but also young isolated neutron stars with proper motions, for instance by tying multiple epochs of deep Hubble Space Telescope (\hst) imaging and data from other observatories to the same \gaia\ reference frame \citep{2022ApJ...933...76D}. This allows us to probe much deeper than \gaia\ alone, revealing fainter, more dust obscured companions, as well as the remnants themselves \citep{2022ApJ...926..121L}, while retaining both the accuracy and precision of \gaia\ astrometry. The advent of \jwst\ provides another opportunity in this respect, allowing searches even deeper into the Galaxy. This raises the prospect of statistically significant samples of unbound companions. Upcoming multi-object spectrographs and integral field units such as WEAVE \citep[WHT Enhanced Area Velocity Explorer,][]{2014SPIE.9147E..0LD} and 4MOST \citep[4-metre Multi-Object Spectrograph Telescope,][]{2014SPIE.9147E..0MD} also offer new opportunities to characterise candidates in the vicinity of supernova remnants and young neutron stars, through the identification of stars with high radial velocities, high rotational velocities or unusual chemical abundances, which may be indicative of past binary interactions \citep[e.g.][]{2014A&A...565A..90C,2022arXiv220803999G}.

In this paper, we investigate the scope for expanding the sample of unbound supernova companions in the Milky Way to a size capable of testing population synthesis predictions for the binary progenitors of supernovae and neutron star natal kicks. Using the Binary Population and Spectral Synthesis \citep[BPASS][]{2017PASA...34...58E,2018MNRAS.479...75S} stellar evolution models, we produce photometric and kinematic predictions for the outcomes of supernovae in binaries, and compare with current observational constraints and the capabilities of contemporary and upcoming facilities. Throughout, uncertainties are at the 1$\sigma$ level. Magnitudes are quoted in the AB system \citep{1983ApJ...266..713O}, with the exception of \gaia\ magnitudes, which are given in the Vega system for ease of comparison with \gaia\ results quoted elsewhere in the literature.

\section{Supernovae in binaries}\label{sec:explode}
\subsection{Binary population synthesis models}\label{sec:bpass}
We use the stellar evolution models of BPASS v2.2.1 \citep[Binary Population and Spectral Synthesis,][]{2017PASA...34...58E,2018MNRAS.479...75S}. These consist of binary and single star models, weighted in proportion according to the observed distribution of stellar and binary parameters including the binary fraction, initial mass function, orbital periods and mass ratios \citep{2017ApJS..230...15M}. Throughout, we use a metal mass fraction of $Z=0.020$, which corresponds to approximately Solar metallicity.

For each model, we first determine whether the star will go supernova. Our supernova selection criteria are as follows \citep[and follow standard BPASS criteria, see e.g.][]{2017PASA...34...58E,2019MNRAS.482..870E,2020MNRAS.491.3479C,2022MNRAS.514.1315B}. We require a final total mass of $>$1.5\,M$_{\odot}$, a carbon-oxygen core mass of $>$1.38\,M$_{\odot}$, and at least 0.1\,M$_{\odot}$ of oxygen and neon in the core, to ensure that burning has progressed past the point of white dwarf formation. Finally, we require a neutron star remnant, rather than a black hole. This is because we are interested in companions associated with successful, rather than failed, supernovae, and the number of successful supernova producing black holes is thought to be small \citep{2020ApJ...890...51E}. The remnant mass is calculated by assuming a typical supernova energy of 10$^{51}$\,erg. Thin layers are removed from the surface of the star repeatedly, with the binding energy required to remove each layer subtracted from the supernova energy budget each time, until none is left. The total mass removed at the end of this process is considered the ejecta mass, the remainder is the remnant mass \citep[for full details, see][]{2004MNRAS.353...87E}. The upper mass limit for neutron star versus black hole formation is at present unclear; the most massive neutron stars observed may have masses as high as 2.4--2.7\,M$_{\odot}$ \citep{2002A&A...392..909C,2008ApJ...675..670F} with theoretical predictions up to $\sim$3\,M$_{\odot}$ \citep{2012ARNPS..62..485L}. We adopt $1.4<M/M_{\odot}<3.0$ as the neutron star mass range for consistency with other BPASS outputs \citep{2017PASA...34...58E}, and note that the BPASS remnant mass distribution is strongly biased towards the lower masses in this range, so the precise choice of upper mass cut will not be a dominant factor influencing our results.

We next split the exploding models by the spectral type of supernova they are predicted to produce, to see if our predictions vary significantly between hydrogen-rich type II and stripped envelope type Ibc supernovae. For type II supernovae, we require a minimum total hydrogen mass of $10^{-3}$\,M$_{\odot}$ \citep[at the lower end of literature estimates,][]{2011MNRAS.414.2985D,2022MNRAS.511..691G}, and type Ibc supernovae occur when there is less hydrogen remaining. We find, by summing the weights of models which go SN, that approximately 25 per cent of SNe arise from single stars (initially single or merged binaries), 25 percent from secondary stars (bound and unbound), 5 per cent from primaries which remain bound post-SN and the remaining 45 per cent from primaries which unbind the binary. These numbers are obtained from the publicly available BPASS v2.2.1 models \citep{2018MNRAS.479...75S}\footnote{\url{https://bpass.auckland.ac.nz/}}, assuming the fiducial broken power-law initial mass function (IMF) with masses in Solar units as follows,
\begin{equation}
    N(M<M_{\rm max}) \propto \int_{0.1}^{0.5} M^{-1.30} dM  +  0.5^{-1.30} \int_{0.5}^{M_{\rm max}} M^{-2.35} dM
\end{equation}
and the \citet{2017ApJS..230...15M} binary parameter distributions (see also the following analysis, which reproduces these numbers). Investigations of the effect of varying the input binary parameter distributions in BPASS were carried out by \citet{2020MNRAS.495.4605S} and \citet{2022MNRAS.513.3550C}. 

The results of any population synthesis, in addition to the input initial binary parameter distributions, are determined by the subsequent modelling assumptions. BPASS assumes, if interactions occur, that the accretion rate of the secondary is limited by the thermal timescale. Any mass accreted above this limit is instead lost from the system; to calculate the angular momentum lost (and hence the orbital evolution) this mass removal is assumed to occur in spherically symmetric shells around the donor. The treatment of the binary angular momentum then closely follows \citet{2000A&A...360.1011N} and \citet[][the $\gamma$ formalism]{2005MNRAS.356..753N}. If common envelope evolution (CEE) occurs, BPASS has the benefit of modelling the detailed internal structure of the primary. However, CEE timescales are artificially long since a maximum mass-loss rate of 0.1\,M$_{\odot}$\,yr$^{-1}$ is imposed - during CEE only - due to numerical constraints (i.e. the total mass loss is the same, but the time taken to lose this mass is lengthened). Since the 0.1\,M$_{\odot}$\,yr$^{-1}$ limit only applies {\it during} CEE, it does not change the number of systems entering this phase. When expressed in terms of the $\alpha$-$\lambda$ formulation, BPASS CEE gives $\alpha \lambda$ values in the range $2-30$ \citep{2017PASA...34...58E,2023MNRAS.520.5724B}. 

Stellar rotation is tracked, but has no impact on the models other than for deciding whether a star undergoes quasi-homogeneous evolution, and tidal interactions are not implemented in current releases (although synchronisation is assumed to occur once Roche lobe overflow occurs). Wind-driven mass loss rates follow \citet{1988A&AS...72..259D}, except for OB stars which use the rates of \citet{2001A&A...369..574V}, and Wolf-Rayet winds which are based on \citet{2000A&A...360..227N}. These processes (and more) are described in full by \citet{2017PASA...34...58E}.

\subsection{Natal kicks for supernovae in binaries}
BPASS makes use of the \citet{1998A&A...330.1047T} and \citet{1999MNRAS.310.1165T} model for calculating the outcomes of supernovae in binaries. Given the pre-supernova masses of the two components, their separation, the ejecta mass of the exploding star and the mass/kick velocity of the remnant, we can determine (i) whether the system is unbound and (ii) the resulting velocity vectors of either the bound secondary-remnant system, or the unbound neutron star and secondary. The inputs to the calculations are taken from the final lines of standard BPASS binary models where the primary goes supernova. The sum of $M_{\rm ejecta}$ and $M_{\rm remnant}$ is the final primary mass pre-supernova, and circular orbits are assumed. Another parameter in the model is the radius of the secondary, but this does not come into play, as we assume that the impact of the ejected mass on the secondary is negligible \citep[measurable effects are expected to be rare,][]{2015A&A...584A..11L,2018ApJ...864..119H,2021MNRAS.505.2485O}.

A final input is the kick velocity distribution of natal neutron stars. These kicks are expected to arise due to asymmetries in the core-collapse process \citep{2017ApJ...837...84J}, and we consider four possible kick distributions as follows:
\begin{enumerate}
    \item The BPASS fiducial model is a Maxwell-Boltzmann distribution with $\sigma = 265$\,km\,s$^{-1}$ \citep{2005MNRAS.360..974H}. For each primary star, we draw 1000 remnant velocities from this distribution and select a random kick direction for each, in $\theta$ (azimuthal direction in the orbital plane) and $\phi$ (angle out of the orbital plane). We draw $\theta$ values from the range $0 < \theta < \pi$, weighted by $sin(\theta)$, and $\phi$ from $-\pi < \phi < \pi$, to ensure kick direction isotropy. From a remnant velocity, the other parameters provided above, and momentum conservation, the fate of the binary and the velocity of the components can be determined.

    \item In addition to the \citet{2005MNRAS.360..974H} pulsar velocity distribution, we also run the code across all binary models using the \citet{2017A&A...608A..57V} bimodal velocity distribution of young pulsars, to investigate the effect of assuming different kick velocity distributions.

    \item We also explore the kick model of \citet{2016MNRAS.461.3747B}, \citet{2018MNRAS.480.5657B}, and \citet{2022arXiv220802407R}, which relates the kick velocity $V_{k}$ to the ejecta mass via,
    \begin{equation}
        V_{\rm k} = \alpha\Bigl(\frac{M_{\rm ejecta}}{M_{\rm remnant}}\Bigr) + \beta\Bigl(\frac{{1.4}}{M_{\rm remnant}}\Bigr) 
    \end{equation}
    where masses are in Solar units. The current best-fit parameter values are $\alpha=115\pm35$\,km\,s$^{-1}$ and $\beta=10\pm10$\,km\,s$^{-1}$, we adopt the central values of 115\,km\,s$^{-1}$ and 10\,km\,s$^{-1}$ \citep{2022arXiv220802407R}.

    \item Finally, we produce predictions for the case where core-collapse is perfectly symmetric, and the natal neutron star does not receive a kick. The recoil velocity of the binary due to symmetric mass loss alone is known as the Blaauw kick \citep{1961BAN....15..265B,1961BAN....15..291B}, or mass-loss kick, and is included in the \citet{1998A&A...330.1047T} model. Velocities arising from Blaauw kicks alone are given by,
    \begin{equation}
        V_{\rm Blaauw} = \frac{M_{\rm ejecta}}{M'} \frac{M_{2}}{M} \times \sqrt{\frac{GM}{a}} 
    \end{equation}
    where M is the total mass of the binary before the supernova, and M$^{'}$ is the total binary mass after the supernova, such that $M^{'} = M - M_{\rm ejecta}$. The secondary mass is $M_{2}$, while $a$ is the semi-major axis of the pre-supernova orbit. We verified that setting natal kick magnitudes to 0 in the \citet{1998A&A...330.1047T} model reproduces the Blaauw distribution, as expected when only a mass-loss recoil occurs in the system.
\end{enumerate}

\section{Results: kinematics}
Sections \ref{sec:unbounds}, \ref{sec:remnants} and \ref{sec:bounds} describe results for the Hobbs and Blaauw kicks, results for the Verbunt \citep{2017A&A...608A..57V} and Bray \citep{2016MNRAS.461.3747B} kicks are discussed in Section \ref{sec:altkick} and shown in Appendix \ref{sec:apx1}.

\subsection{Unbound secondaries}\label{sec:unbounds}
\begin{figure*}
	\includegraphics[width=0.99\textwidth]{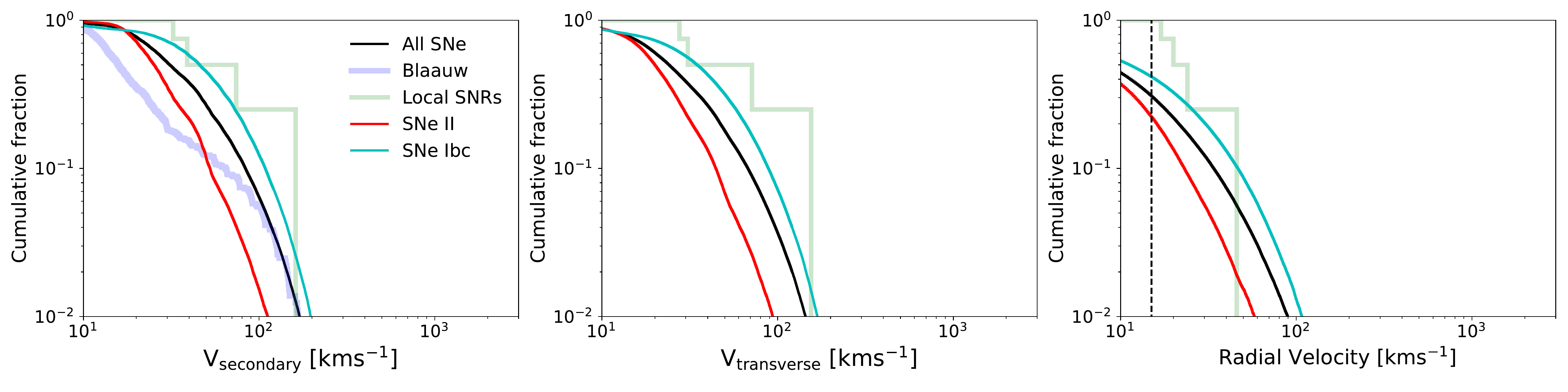}
    \caption{Cumulative distributions of predicted unbound secondary star velocities. Left: total velocities for secondaries ejected by type II, type Ibc and all supernovae (thin red, cyan and black lines). The results from assuming only Blaauw kicks are shown by the thick blue line. Middle: transverse velocities, assuming random viewing angles, observable as proper motions. Right: radial velocities, again assuming random orientations with respect to the line of sight. The $R \sim 20000$ minimum velocity measurable with with WEAVE and 4MOST is indicated by a vertical dashed line. Four candidates for local ejected supernova companions are shown in green \citep[][see Sec. \ref{sec:snrs}]{2017A&A...606A..14B,2019ApJ...871...92F}.}
    \label{fig:unbounds}
\end{figure*}

Figure \ref{fig:unbounds} shows the kinematic results for the case where the secondary is ejected. The velocities are calculated according to equations 54--56 of \citet{1998A&A...330.1047T}. The three panels show the total 3D velocity, projected 2D velocities (observable as proper motions) and projected 1D (radial) velocities. For the 2D and 1D projections, random viewing angles are assumed - each primary model is exploded 1000 times, and each time a random kick magnitude, kick direction and viewing angle are chosen. An indicative minimum measurable radial velocity of 15\,km\,s$^{-1}$ is shown on the 1D panel, corresponding to a spectral resolution of $R=20000$, comparable to the capabilities of upcoming multi-object spectrographs 4MOST \citep{2014SPIE.9147E..0MD} and WEAVE \citep{2014SPIE.9147E..0LD}. Velocity distributions for secondaries ejected by a type II, type Ibc, and any type of supernova are shown. Secondaries ejected by stripped envelope supernovae are typically faster moving. We find that binaries producing type Ibc events have mean pre-supernova orbital separations of 174\,R$_{\odot}$, compared with $\sim$1000\,R$_{\odot}$ for type II events. Therefore, the higher velocities of ejected Ibc companions is due to tighter pre-supernova orbits. This is symptomatic of envelope stripping playing an important role in producing Ibc events - tighter orbits are more likely to produce stripped envelope primary supernova {\it and} higher secondary velocities, if unbound. This correlation exists because $\sim$75 per cent of the high-mass BPASS primaries which undergo Roche lobe overflow progress to common envelope evolution \citep[see e.g. Figure 1 of][]{2017PASA...34...58E}, tightening the orbits. Consequently, $\sim$15 per cent of BPASS binaries with a primary ZAMS mass greater than 8\,M$_{\odot}$ eventually merge. 

The stability of mass transfer (dependent on how the radius varies with mass loss) is not approximated in BPASS \citep[e.g. with a $\zeta$ factor,][]{2008ApJS..174..223B}, but is instead determined by re-solving the stellar structure equations with the new mass at each model time-step. Combined with the \citet{2017ApJS..230...15M} input binary parameter distributions, this results in a high proportion of systems undergoing CEE, despite recent demonstrations that detailed models typically have more stable mass transfer than rapid codes \citep[e.g.][]{2017MNRAS.465.2092P,2021A&A...650A.107M,2023A&A...669A..45T}. Although the fraction of systems progressing from RLOF to CEE or merger is high, it is broadly in line with the observed value of $\sim$80 per cent for O-stars \citep[e.g.][]{2012Sci...337..444S}. In any case, the fraction of massive binaries progressing to CEE is critical to our results, as it determines the final orbital periods and hence secondary velocities if unbound. 

Also shown on the first panel is the velocity distribution assuming symmetric supernovae (all supernova types, Blaauw kicks only). In order to unbind through the Blaauw mechanism, more than half of the pre-supernova binary mass has to be ejected \citep{1961BAN....15..265B}, making unbinding events through this process rarer. The final velocities of unbound secondaries are between 0.5-1.0 of their pre-SN orbital velocity with respect to the centre of mass, where the reduction occurs due to the climb out of the potential \citep[][]{1961BAN....15..265B,1961BAN....15..291B}. Also shown are measurements of four strong supernova runaway candidates, identified by \citet{2015MNRAS.448.3196D} and \citet{2017A&A...606A..14B}. These are part of a sample of 13 supernova remnants (SNRs) which we use as a comparison dataset in Section \ref{sec:snrs}, where these candidates are discussed further. 

As a point of comparison with other population synthesis predictions, we find that the ratio of runaways exceeding 30\,km\,s$^{-1}$ to those exceeding 60\,km\,s$^{-1}$ is $\sim$3 for high-mass ($>$7.5\,M$_{\odot}$) stars. This compares with the fiducial \citet[][{\sc binary$\_$c} models]{2019A&A...624A..66R} value of $\sim$25 for this ratio. We note that BPASS v2 predicts faster unbound velocities in general - we find a median velocity for all unbound companions of 35.6\,km\,s$^{-1}$ (compared with 10.4\,km\,s$^{-1}$), and that $\sim$40 per cent of OB stars meet the runaway 30\,km\,s$^{-1}$ threshold, compared with a few per cent as previously predicted \citep{1997A&A...318..812D,2011MNRAS.414.3501E,2019A&A...624A..66R,2020MNRAS.497.5344E}. The observed fraction of runaways in the observed OB population, which includes some dynamical contribution, is closer to 10--20 per cent \citep[][]{1961BAN....15..265B,1987ApJS...64..545G,1991AJ....102..333S,2005A&A...437..247D,2011MNRAS.410..190T,2018A&A...616A.149M}. 

The input binary parameters are likely driving much of this difference - for instance, \citet{2019A&A...624A..66R} assumes a flat mass ratio distribution, whereas BPASS v2 implements the results of \citet{2017ApJS..230...15M} which can have mass ratio distributions with slopes as a steep as -2 at some orbital separations, with a excess twin fraction at short periods on top of this. We find different results with respect to BPASS v1 \citep[][such as a much higher OB runaway fraction in this work]{2011MNRAS.414.3501E}, which assumed a flat initial period distribution. However, other factors such as different treatments of the mass transfer efficiency, common envelope evolution and envelope stripping will also play a role in creating the discrepancy with other codes (see Section \ref{sec:bpass} and above). In particular, the detailed treatment of mass transfer stability and the common envelope phase are critical in determining the final orbital period distribution, and hence the velocities of unbound secondaries.

\begin{figure*}
	\includegraphics[width=0.99\textwidth]{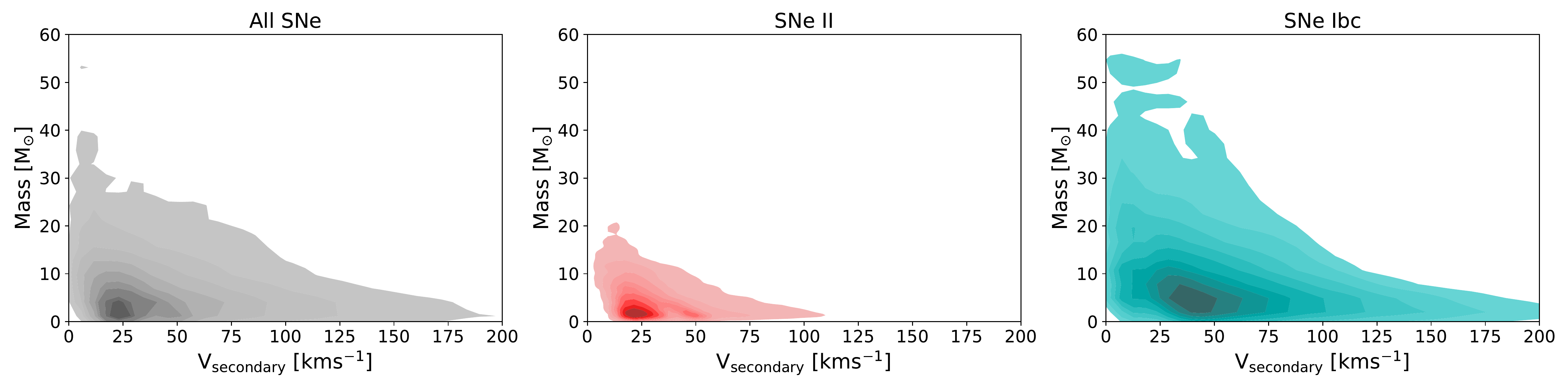}
    \caption{Masses of the secondary stars ejected by the supernovae of primaries, versus their total 3D velocities, assuming \citet{2005MNRAS.360..974H} natal kicks. Left: secondaries ejected by all SNe. Middle: secondaries ejected by SNe type II. Right: secondaries ejected by SNe type Ibc. Secondaries ejected by stripped envelope SNe are more massive and faster moving then those ejected by type II SNe. Ten contour levels are shown; the darkest shade encloses 10 per cent of the population, one shade lighter encloses 20 per cent, and so forth. The lightest colour plotted encloses 90 per cent, the remaining 10 per cent of ejected secondaries occur outside the shaded areas.}
    \label{fig:massvel}
\end{figure*}

\begin{figure*}
	\includegraphics[width=0.99\textwidth]{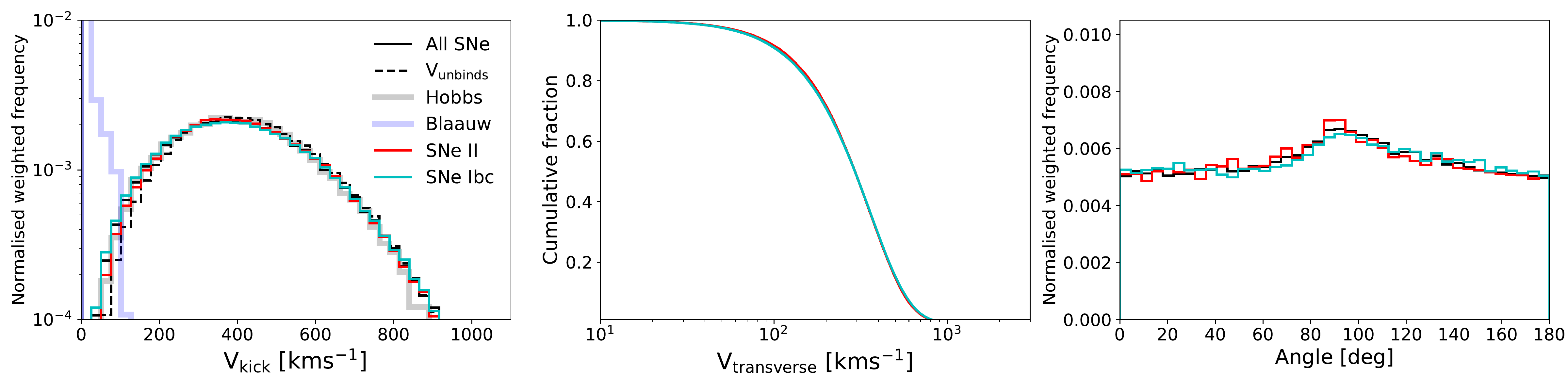}
    \caption{Kinematic results for primary neutron star remnants, in the case where the binary unbinds. Left: the total 3D velocity distribution of ejected neutron stars. The dashed line is the subset of input \citet{2005MNRAS.360..974H} natal kicks $V_{\rm kick}$ which result in the binary being unbound. The solid black, red and cyan lines are the final unbound neutron star velocities from different classes of supernovae. The final velocity distributions are similar to the input kicks, demonstrating that the Blaauw kick (thick blue line) is a small contribution. The final velocities move even closer to the input kicks than $V_{\rm unbinds}$ due to the vectorial addition of the pre-supernova orbital velocities. Middle: transverse neutron star velocities, assuming random orientations, measurable as proper motions. Right: the angle between the neutron star and unbound secondary velocity vectors. This distribution is close to flat, again indicating that natal kicks dominate. Randomly drawn isotropic kick directions, circular orbits and random viewing angles are assumed.}
    \label{fig:remnants}
\end{figure*}

\begin{figure*}
	\includegraphics[width=0.99\textwidth]{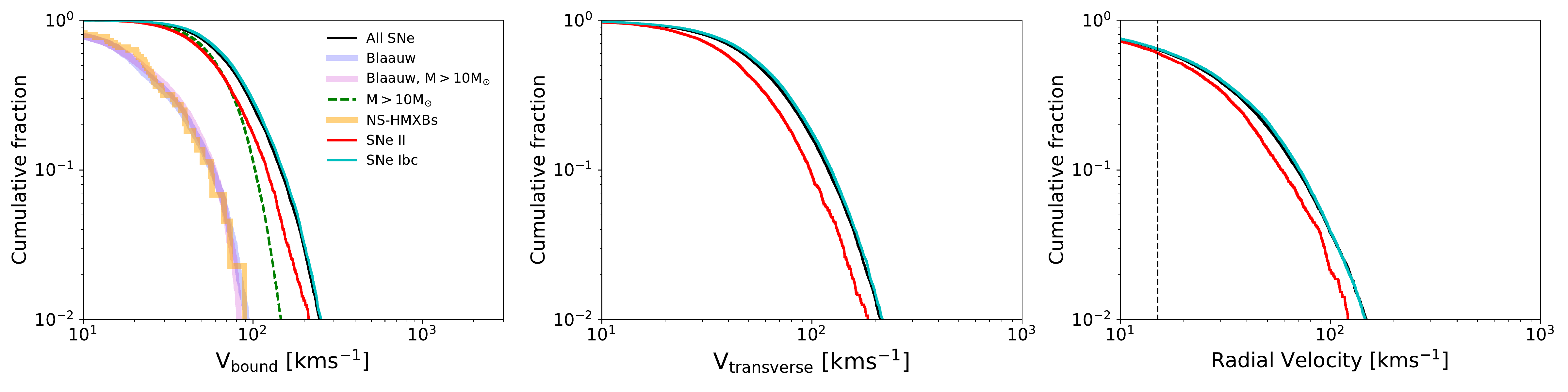}
   \caption{As in Figure \ref{fig:unbounds}, but for bound neutron star/non-degenerate systemic velocities. Blaauw kicks are shown by the thick blue line, Blaauw kicks for systems with a companion greater than 10\,M$_{\odot}$ by the thick pink line. Results from the inclusion of neutron star natal kicks are given by the red, cyan and black lines. Kicks for bound systems (all supernovae) with a companion and asymmetric natal kicks are shown by the dashed green line. The NS-HMXB peculiar velocities of \citet{2022arXiv220603904F} are shown in yellow. From left to right we show the total (3D) velocity, the project (2D) velocities, and finally the radial (1D) velocity component. The 2D and 1D velocities are derived from the 3D distribution by assuming random orientations with respect to the line of sight. The vertical dashed line on the RV panel represents the minimum radial velocity measurable with WEAVE or 4MOST (assuming $R \sim 20000$).}
    \label{fig:bounds}
\end{figure*}

\subsection{Unbound neutron stars}\label{sec:remnants}
Figure \ref{fig:remnants} shows results for unbound neutron stars, when asymmetric kicks and the \cite{2005MNRAS.360..974H} distribution are assumed. Equations 51--53 of \citet{1998A&A...330.1047T} are used to derive these velocities. The first and second panels are again the total velocities and projected transverse velocities. The final neutron star velocities in the Blaauw case are small, because in the case the binary unbinds, the remnant carries on with the initial orbital velocity of the primary v$_{1}$ \citep[and at infinity has a velocity 0.5-1.0\,v$_{1}$,][]{1961BAN....15..265B,1961BAN....15..291B}.

When asymmetric kicks are implemented, the remnant velocities closely follow the input natal kick distributions. This implies that asymmetric kicks dominate over the recoil (Blaauw) kick, which is consistent with the small (tens of km\,s$^{-1}$) Blaauw kicks shown. Binaries are more likely to remain bound when the kick $V_{\rm kick}$ is small, the subset of input kicks which result in the binary becoming unbound are denoted by $V_{\rm unbinds}$ in Figure \ref{fig:remnants}. This distribution is preferentially missing lower velocities with respect to the input kicks - the distribution of input kicks which result in the system remaining bound peaks at $\sim$200\,km\,s$^{-1}$ with a standard deviation of 130\,km\,s$^{-1}$. However, the final unbound neutron star velocity distributions (all SNe, SNe II, SNe Ibc) are slightly broader than $V_{\rm unbinds}$. This is because these final velocities incorporate the pre-supernova orbital velocity, which can be oriented in any direction with respect to the isotropic kicks. For example, in some cases, the kick and orbital velocity can act in opposite directions, pushing the lower end of the distribution down \citep{1998A&A...330.1047T}. The mean primary velocity pre-supernova is 8.4\,km\,s$^{-1}$, with a tail reaching velocities of hundreds of km\,s$^{-1}$. 

In summary, because only $\sim$10 per cent of binaries remain bound, the difference between the input kicks $V_{\rm kick}$ and the $V_{\rm unbinds}$ distribution is small. The difference between the input kicks and the final unbound velocities is made smaller still by adding the broadening effect of the pre-supernova orbital velocities. The net result is that the final unbound neutron star velocities closely match the input kick distribution. This result has been noted before \citep[e.g.][]{2011MNRAS.414.3501E,2021MNRAS.508.3345I} and justifies the assumption, made widely in the literature \citep[e.g.][]{2005MNRAS.360..974H,2016MNRAS.461.3747B,2017A&A...608A..57V,2020MNRAS.494.3663I,2022arXiv220909252K}, that isolated neutron star velocity distributions inferred from observations are broadly representative of the initial natal kicks. 

In the final panel of Figure \ref{fig:remnants}, we show the projected angle between the remnant and walkaway/runaway proper motion vectors. The distribution shows a slight peak around $\sim$90\,degrees. This is because the secondary motion post-primary supernova, when unbound, is primarily along the direction it was travelling pre-primary supernova. The neutron star velocities, on the other hand, are much higher and are close to being isotropically distributed. The combination of isotropic neutron star velocities and slower secondary velocities being biased in one direction produces the peak at $\sim$90\,degrees, which we verified by artificially increasing the secondary velocities in this direction, noting that this increases the strength of the peak. The right panel of Figure \ref{fig:remnants} shows probability density, such that area under each line is unity. Since 0.005$\times$180$=$0.9, there is a small (but not negligible) excess probability of finding angles in the range $\sim$60--120 degrees.

\subsection{Bound systems}\label{sec:bounds}
Although not the focus of this paper, as a separate observational test, we show in Figure \ref{fig:bounds} kinematic results for the case that the binary remains bound. The systemic velocities of bound systems after an asymmetric supernova are calculated using equation 1 of \citet{1999MNRAS.310.1165T}. As in Figure \ref{fig:unbounds}, the three panels show 3D, 2D and 1D velocities, where random viewing angles are assumed for the latter two, and Blaauw kicks are also shown on the first panel. As an observational comparison, we plot the peculiar velocities of neutron star high-mass X-ray binaries in \gaia\ \citep[NS-HMXBs,][]{2022arXiv220603904F}, which have been corrected for the Solar motion and their local standard of rest, and include both proper motion and radial velocity constraints. 

Compared with the predictions for asymmetric supernovae, the NS-HMXBs are strongly biased to lower velocities. To check whether this arises due to the high donor masses in NS-HMXBs, we also plot the predicted distribution (for all supernovae) where the surviving bound companion has a mass of at least 10\,M$_{\odot}$. This is the minimum donor mass in the NS-HMXB sample, and it shifts the predicted distribution to the left as expected - the binary has more mass for the same neutron star natal kicks. An Andersen-Darling test between the NS-HMXBs and asymmetric kicks with $>$10\,M$_{\odot}$ gives a $p$-value  of 0.001. We note that the NS-HMXB velocities are consistent with Blaauw kicks: an Andersen-Darling (AD) test between the NS-HMXBs and Blaauw kicks with a companion $>$10\,M$_{\odot}$ yields 0.23, failing to reject the null hypothesis that these distributions are consistent (at the 2$\sigma$ level). Although Blaauw kicks are an apparently good fit, it is important to realise that asymmetric kicks in massive binaries - where $M_{2}>>M_{\rm remnant}$ - produce Blaauw-like distributions for bound systems, a result which has been previously noted \citep[e.g.][]{2011A&A...529A..14G}. Although we are unable to reproduce the NS-HMXB distribution with any choice of mass cut, we note that the orbital properties of XRBs are also not representative of the bound population as a whole, and should be considered in order to make this comparison (but that full consideration of these criteria is beyond the scope of this paper).

\begin{figure*}
    \includegraphics[width=0.99\textwidth]{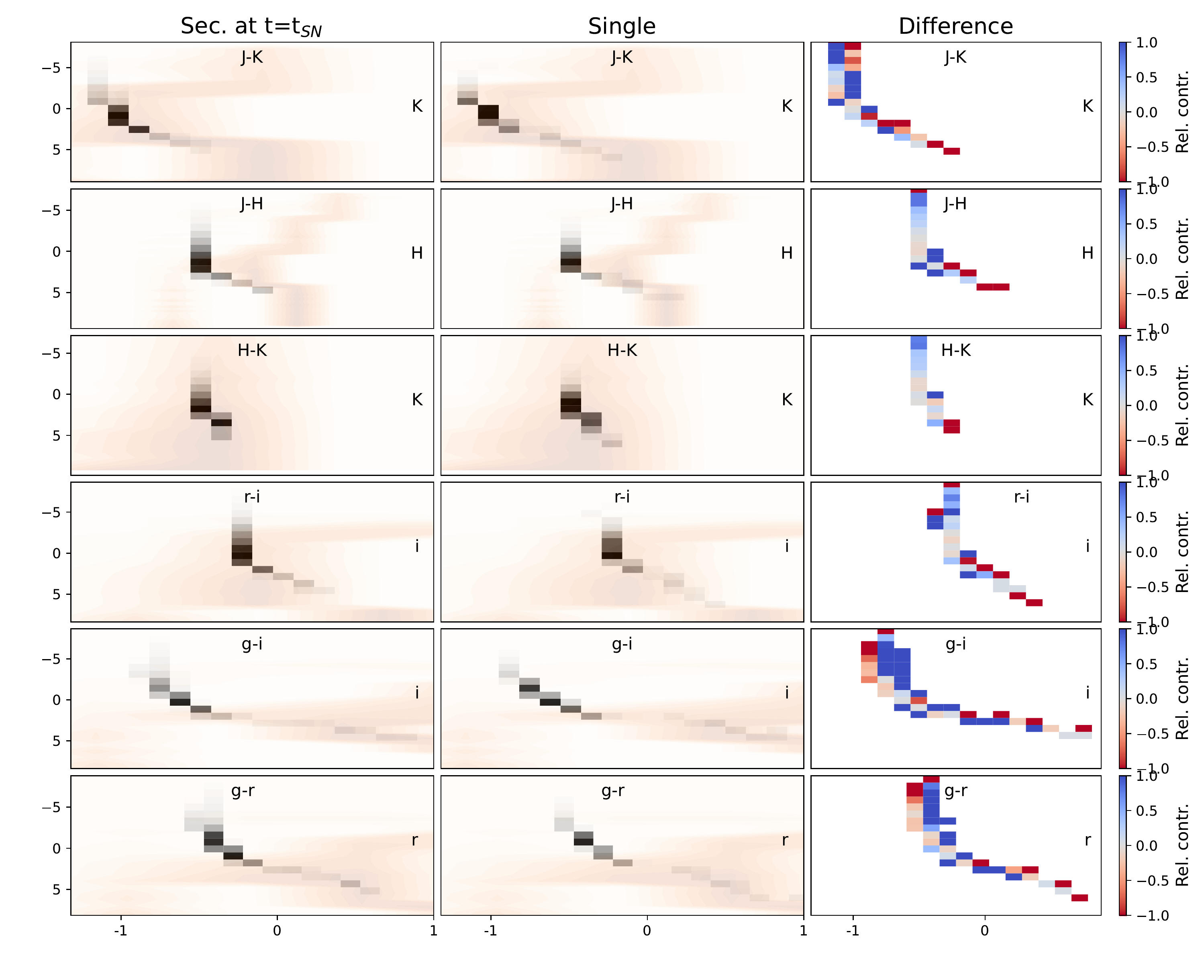}
    \caption{Colour--magnitude diagrams (CMDs) for unbound secondary stars (from all supernovae). The axes labels are on the inside (upper and right) of each plot. Left column: photometry for all unbound companions (greyscale histogram), at the time of primary supernovae, $t_{\rm SN}$. Middle: photometry for single stars with the same ZAMS mass as the secondaries in the left panel, at age $t \sim t_{\rm SN}$. We have applied the binary model weighting to the equivalent single star models (rather than their own IMF weighting) to ensure fair comparison. These results therefore represent the evolution of the secondary stars without any binary evolution assumptions. Right: The relative contribution in each CMD cell from unbound stars at time $t=t_{\rm SN}$ and their single star equivalent photometry, defined as (N$_{\rm SN}$-N$_{\rm sin}$)/(N$_{\rm SN}$+N$_{\rm sin}$). Bluer (more positive) values indicate a higher contribution to that cell from secondary stars at $t=t_{\rm SN}$, where a value of 1 indicates that no single stars contribute. Redder (more negative) values can reach $-$1, corresponding to exclusively single stars (i.e. all secondary stars leave this CMD cell before they are ejected by their companions supernova). The background (orange) shading of the left and middle panels is a simple stellar population of age 10$^{10}$ years \citep[produced with HOKI,][]{2020JOSS....5.1987S} representing the field population of stars in the Galactic disc.}
    \label{fig:photometry_unbounds}
\end{figure*}

\begin{figure*}
    \includegraphics[width=0.99\textwidth]{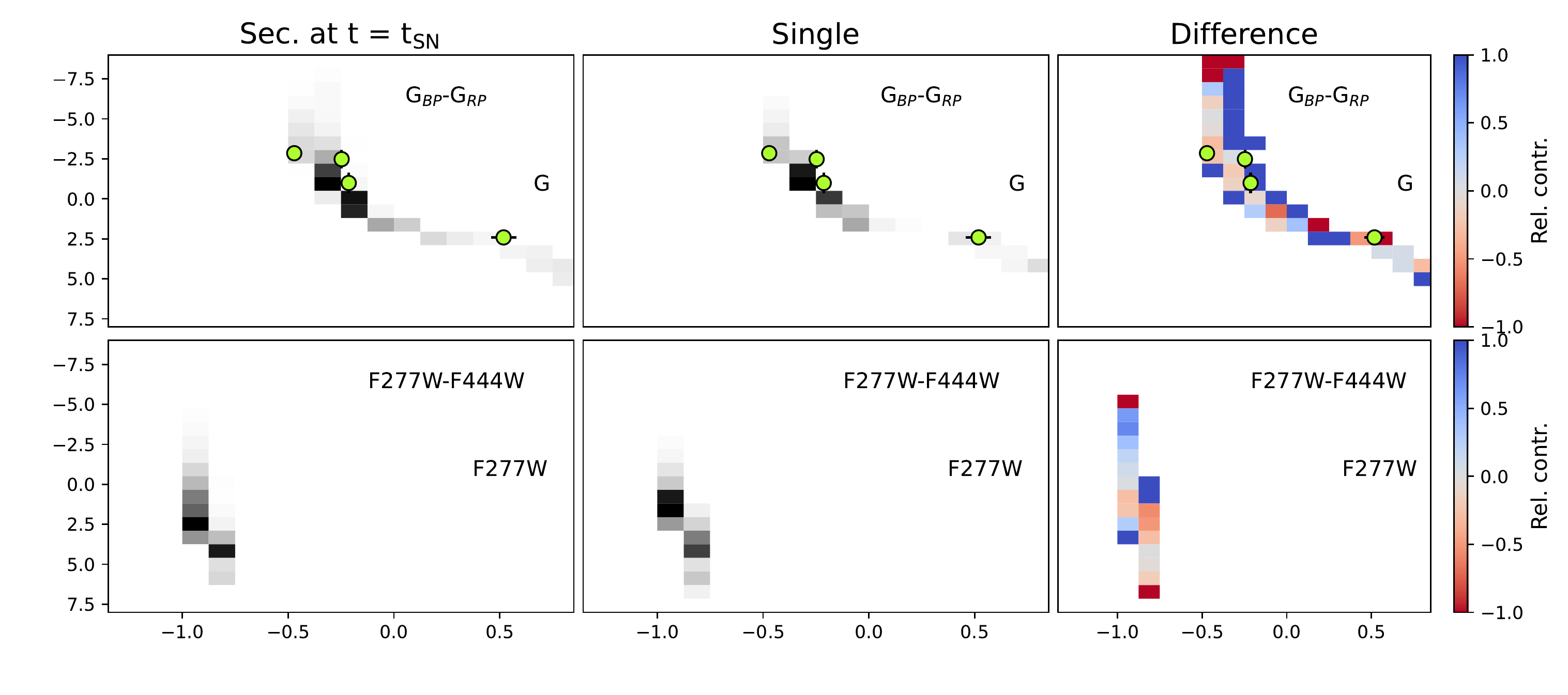}
    \caption{Colour--magnitude diagrams (CMDs) for unbound secondary stars for the (\gaia) G-band vs $G_{\rm BP}$-$G_{\rm RP}$, and F277W vs F277W-F444W (\jwst/NIRCam filters). \jwst\ photometry is given in the AB system, \gaia\ magnitudes are with respect to Vega. As in Figure \ref{fig:photometry_unbounds}, the left and middle panels show the secondaries at the time of primary supernova and equivalent-mass single stars respectively, and the right panels show the relative contribution to each CMD cell from these. Unlike Figure \ref{fig:photometry_unbounds}, field stars are not shown here, as \gaia\ and \jwst\ magnitudes are not standard BPASS outputs available through HOKI. The four candidates for runaways ejected from local SNRs are shown on the \gaia\ panels \citep[][see Sec. \ref{sec:snrs}]{2017A&A...606A..14B,2019ApJ...871...92F}.}
    \label{fig:photometry_gaiajwst}
\end{figure*}

\subsection{Alternative kick prescriptions}\label{sec:altkick}
In each of Figures \ref{fig:unbounds}-\ref{fig:bounds}, the total velocity when Blaauw kicks are assumed is also shown. We find that with Blaauw recoil kicks alone, $\sim$20 per cent of systems unbind upon primary supernova. This compares well with the value of 16 per cent found by \citet{2019A&A...624A..66R}, and is far less than the $\sim$80--90 per cent unbinding fraction if asymmetric remnant kicks are also considered \citep[as previously demonstrated, e.g.,][]{2019A&A...624A..66R}. The resultant remnant velocities are low in the Blaauw case. There are signs of a low-kick velocity subset in the Galactic pulsar population \citep{2017A&A...608A..57V,2021MNRAS.508.3345I}, possible arising from electron-capture supernovae \citep{2019MNRAS.482.2234G,2020ApJ...891..141G}, not considered in this work, or ultra-stripped supernovae \citep{2021ApJ...920L..37W}. If there exists such a low-velocity population, this has important implications for the evolution and merger locations of binary neutron star systems \citep[e.g.][Gaspari et al., in prep]{2021MNRAS.503.5997P}. However, the current focus of this paper is on unbound secondaries. We also run the binary supernova code using the distributions of \citet{2017A&A...608A..57V} and \citet{2022arXiv220802407R} for the input neutron star natal kicks, the results of which are shown in Appendix \ref{sec:apx1}. The bimodal \citet{2017A&A...608A..57V} distribution, and Bray \citep[i.e.][]{2016MNRAS.461.3747B} kicks, include proportionally more slower kicks, albeit still faster than those from the Blaauw mechanism. The bound fraction is increased by $\sim$ 10 per cent with the Verbunt kick prescription, while the adoption of Bray kicks \citep[with the parameters of][]{2022arXiv220802407R} produces even fewer unbound systems - only $\sim$70 per cent unbind, opposed to $\sim$90 per cent with Hobbs \citep[see also][]{2022MNRAS.513.3550C}. 

We also note that when comparing the \citet{2022arXiv220603904F} NS-HMXB sample with the bound system predictions, unreasonably high minimum masses are required for the Verbunt and Bray predictions to achieve consistency - a minimum donor mass of $\sim$40--50\,M$_{\odot}$ is required to return AD-test $p$-values in excess of 0.05 for Verbunt, and no choice of minimum mass produces consistency when the Bray kick is used. However, we again caution that a simple minimum mass cut simplifies the situation, and other criteria (such as the orbital separation) should also be considered to reproduce X-ray binary samples.

\begin{figure*}
	\includegraphics[width=0.99\textwidth]{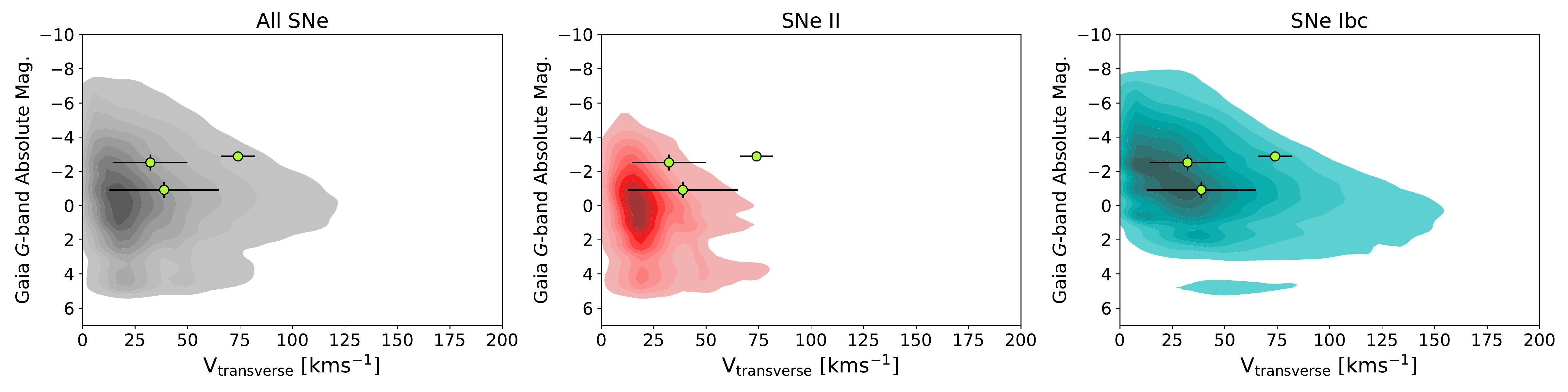}
    \caption{\gaia\ $G$-band absolute magnitudes (Vega) versus 2D projected velocities. The contour levels are as described in Figure \ref{fig:massvel}. The three candidate supernova remnant runaways with well constrained velocities (described in Section \ref{sec:snrs}) are also shown.}
    \label{fig:magvel}
\end{figure*}

\section{Results: photometric properties}
We now examine the predicted absolute magnitudes and intrinsic colours of the unbound secondaries. Figures \ref{fig:photometry_unbounds} and \ref{fig:photometry_gaiajwst} show colour-magnitude diagrams (CMDs) for unbound companions \citep[assuming kicks from][]{2005MNRAS.360..974H}. Each row corresponds to a different colour-magnitude diagram. The first six, in Figure \ref{fig:photometry_unbounds}, are constructed from photometry provided as default BPASS outputs ($g$, $r$, $i$, $J$, $H$ and $K$-bands). Details of how photometry is generated from the stellar atmosphere models and resulting spectral energy distributions are given by \citet{2017PASA...34...58E}.

Photometry is taken from secondary star parameters in the last lines of binary models, before the primary goes supernova \citep[this does not consider the possible effects of supernova ejecta impacting the secondary, e.g.,][ we will return to this point later]{2018ApJ...864..119H}. The rows in Figure \ref{fig:photometry_gaiajwst} are \gaia\ and \jwst\ specific predictions. The \gaia\ photometry is constructed from the $V$ and $I$-bands, following the relations of \citet{2021A&A...649A...3R}, and is provided in the Vega magnitude system (all other photometry is reported in AB magnitudes). The \jwst/NIRCam photometry is calculated by fitting stellar spectra to the $g$, $r$, $i$, $J$, $H$ and $K$ photometry of each model, and applying F277W and F444W filter response curves to the best-fit spectra to obtain magnitudes in these filters. More details on the construction of the \gaia\ and \jwst\ photometry are provided in Appendix \ref{sec:apx2}. 

Figures \ref{fig:photometry_unbounds} and \ref{fig:photometry_gaiajwst} also have three columns: the first shows CMDs for the secondaries at the time of primary supernova. The second shows the photometry of single stars with the same ZAMS mass as the secondaries, also at age $t=t_{\rm SN}$. Therefore, this represents the photometry of the secondaries before any binary evolution specific effects are taken into account. The impact of the assumptions made when modelling these effects (e.g. common envelope evolution, stability of mass transfer etc) can be large \citep[e.g.][]{2022arXiv221111774I}, so the inclusion of equivalent age and ZAMS-mass single star photometry provides a reference point before binary evolution assumptions have taken effect (although we note that these also play a role in deciding which systems are classified as bound/unbound).

The background shading in columns 1 and 2 of Fig. \ref{fig:photometry_unbounds} is the field population of stars, represented by a BPASS simple stellar population (single burst of star-formation) of age 10$^{10}$\,yr \citep[plotted with HOKI,][]{2020JOSS....5.1987S}. The final column shows the relative contribution to each CMD cell from secondaries at $t=t_{\rm SN}$, and at ZAMS. The scale in the final column is defined as (N$_{\rm SN}$-N$_{\rm ZAMS}$)/(N$_{\rm SN}$+N$_{\rm ZAMS}$), such that 1.0 (blue) corresponds to every star in that CMD cell being at $t=t_{\rm SN}$, and -1.0 (red) exclusively at ZAMS. 

From these plots we can make several observations. Recently unbound secondaries are preferentially bluer and more luminous than typical stars in the Galactic disc, and this is true for single stars also. Many systems undergo mass transfer as the primary evolves, increasing the secondary mass. The mean mass of the unbound secondaries at ZAMS is 7.05\,M$_{\odot}$, while at $t=t_{\rm SN}$ it is 7.31\,M$_{\odot}$. Split by the primary supernova classification, the mean ejected secondary masses are 4.10\,M$_{\odot}$ and 4.38\,M$_{\odot}$ for type II events (at ZAMS and $t=t_{\rm SN}$), whereas for type Ibc SNe the mean masses are 10.44\,M$_{\odot}$ and 10.88\,M$_{\odot}$. The preference for supernova companions to also be massive stars is due to the mass ratio distribution disfavouring unequal mass ratios \citep{2017ApJS..230...15M}. Furthermore, type Ibc progenitors have more massive companions even at ZAMS, again due to the mass ratio distribution and because more massive stars are more likely to strip their envelopes through winds. The higher total binary mass of type Ibc progenitor systems also contributes to their pre-supernova orbital velocities being faster, in addition to smaller separations (see Section \ref{sec:unbounds}). The masses of type Ibc companions only increase slightly more than type II companions on average, which might imply that envelope stripping by a companion is a secondary mechanism for producing SNe Ibc at $\sim$Solar metallicities. However, this picture is complicated by the mass transfer efficiency. In BPASS, the secondary accretion rate is limited by the thermal time-scale, with any excess mass transfer above this limit being lost from the system \citep{2017PASA...34...58E}. Given that the binaries producing Ibc events have tighter pre-supernova orbits, we can see that primary envelope stripping is important, even if much of the stripped mass does not accrete onto the companion. Furthermore, stars massive enough to strip their envelopes entirely through winds are expected to produce black holes \citep[possibly with little to no natal kick, and typically no successful supernova,][]{2021A&A...656L..19Z}. Hence, in systems where neutron stars are formed, envelope stripping by a companion is expected to be an important mechanism. The small difference between columns 1 and 2 in Figures \ref{fig:photometry_unbounds} and \ref{fig:photometry_gaiajwst} nonetheless demonstrates that binary evolution effects - at least the ones accounted for in BPASS v2 - are not having a large impact on the photometry of most secondaries. It also shows that the secondaries, being lower mass, evolve on longer time-scales, and will typically still be on the main sequence at the time of primary supernova. 

On the \gaia\ panel in Figure \ref{fig:photometry_gaiajwst}, we have overlaid the extinction-corrected absolute magnitudes and colours of the four runaway candidates from the sample of \citet{2015MNRAS.448.3196D} and \citet{2017A&A...606A..14B} (see Section \ref{sec:snrs}). We have assumed that the companions have the same photometric properties post-primary supernova as immediately before, which may not be the case if it has been inflated due to the impact of supernovae ejecta. \citep{2021MNRAS.505.2485O} run simulations assuming a separation of 40\,AU, this is similar to the pre-SN orbital separations we find in BPASS, but the effect is expected to be short-term, lasting $\sim$10--100 years. This has implications for surviving companion searches in extragalactic supernova on a time-scale of years post-explosion, and for Galactic searches where the remnant is very young. However, for the majority of applications, ejected companions are typically expected to have relaxed by the time of observation. In any case, noticeable effects are expected to be rare \citep{2015A&A...584A..11L,2018ApJ...864..119H,2021MNRAS.505.2485O}.

\begin{figure*}
	\includegraphics[width=0.99\textwidth]{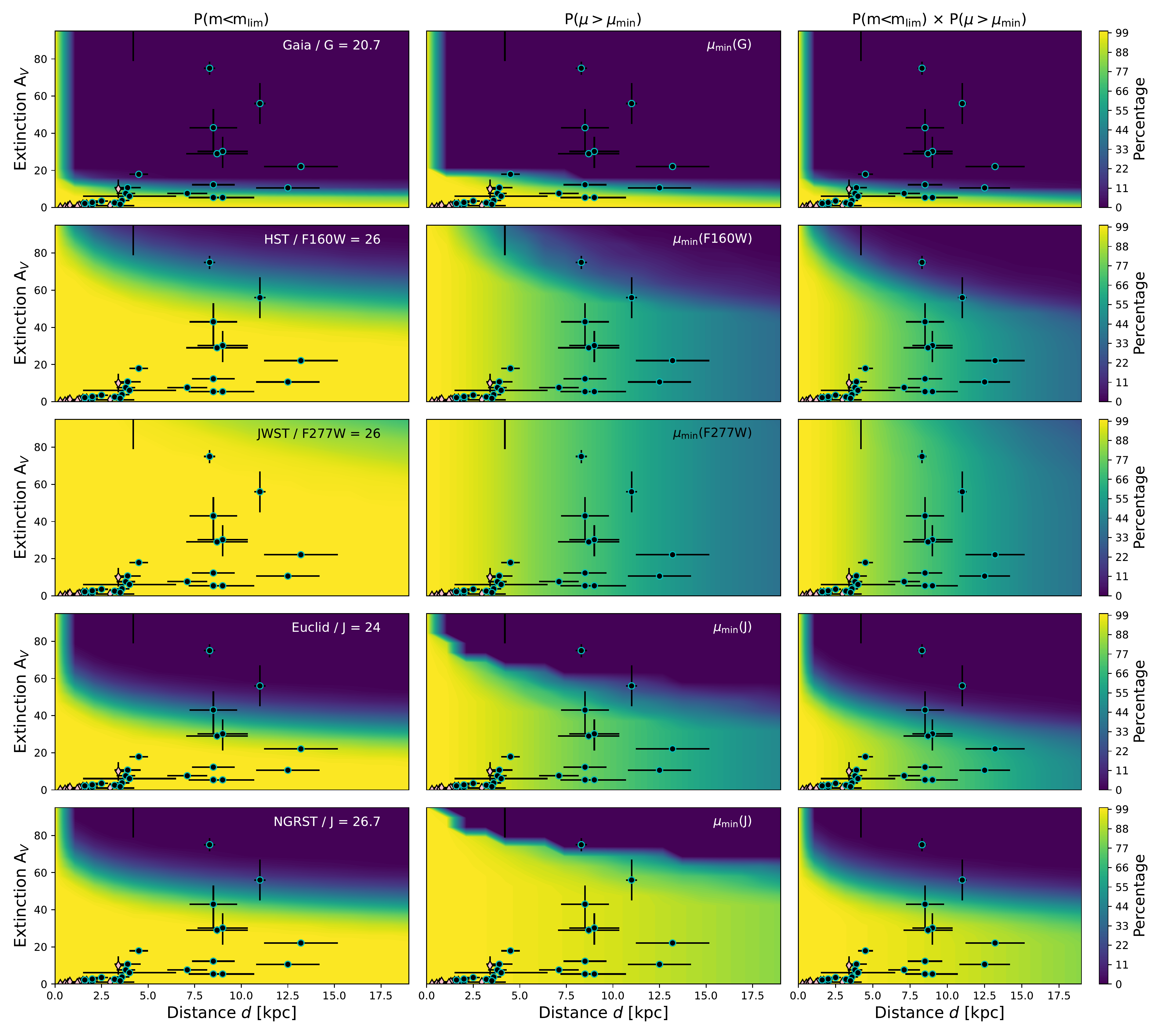}
    \caption{For each column, the shading shows the percentage of the unbound companion distribution which is above the limiting magnitude ($P(m<m_{\rm lim}$), left) and moving faster than the minimum measurable proper motion $\mu_{\rm min}$ ($P(\mu>\mu_{\rm min}$), middle) as a function of distance and extinction $A_{\rm V}$. The final column is the product of the left and middle columns, and represents the fraction of unbound stars which are both detectable {\it and} moving fast enough to have their transverse motion measured. Here we show results for all supernovae using the \citet{2005MNRAS.360..974H} kick. The rows represent different surveys/observatories. Top: predictions for \gaia\ DR3 in the $G$-band (Vega). In this case, we have varied $\mu_{\rm min}$ with magnitude - $\mu_{\rm min}$ at each $d$ and $A_{\rm V}$ corresponds to $\mu_{\rm min}$ for the median magnitude of stars which are above the fiducial limiting magnitude \citep[G=20.7,][]{2021A&A...649A...6G}. The next two rows are \hst\ and \jwst\ examples, assuming F160W ($\sim$H-band) and F277W limiting magnitudes of 26 respectively, with a 5 year baseline, 3 exposures per epoch and $\mu_{\rm min}$ depending on magnitude as described in Section \ref{sec:prob}. Bottom two rows: {\em Euclid} with a $J=24$ limiting magnitude and \ngrst\ with  $J<26.7$, with 100 exposures in each case. For context, the 13 SNRs searched by \citet{2017A&A...606A..14B} and \citet{2019ApJ...871...92F} are show as diamonds (pink), and Galactic magnetars as circles \citep[black/cyan,][]{2014ApJS..212....6O,2022MNRAS.513.3550C}. Note that several SNR points are obscured by magnetar points on this scale.}
    \label{fig:gaiaconstrain}
\end{figure*}

\section{Applications}
\subsection{Probability of detection}\label{sec:prob}
In the context of proper motion searches for runaways and walkaways, the joint distribution of magnitudes and velocities dictates the probability of detection, which also depends on the distance, extinction and proper motion sensitivity. Figure \ref{fig:magvel} shows how the \gaia\ $G$-band absolute magnitudes and projected transverse velocities of ejected secondaries are related, for all supernovae, plus type II and type Ibc events separately. As noted in the previous sections, stars ejected by type Ibc events are more massive and faster moving. 

Overlaid on Figure \ref{fig:magvel} are the magnitudes and velocities of the four strong SNR runaways candidates, discussed in Section \ref{sec:snrs}. For any given proper motion survey, or pair of observations between which a proper motion can be measured, four factors determine whether a runaway or walkaway will be detectable {\it and} moving fast enough for the motion to be measurable. Assuming a fixed set of predictions for unbound companion magnitudes and velocities, these parameters are (i) the imaging depth, (ii) the distance to the SNR or young neutron star, (iii) the extinction along this line of sight and (iv) the minimum measurable proper motion (as a function of magnitude). In Figure \ref{fig:gaiaconstrain}, we demonstrate this with plots of extinction versus distance. The rows represent different observatories. For the first column, the colour scale represents the percentage of the unbound companion magnitude distribution which is detectable for the stated limiting magnitude and filter. The middle column shows the percentage of the transverse velocity distribution detectable, for a given minimum measurable proper motion (the values for the transverse velocities are taken from the middle panel of Figure \ref{fig:unbounds}, where for each model 1000 random kick magnitudes, kick directions and viewing angles to the plane of the binary are drawn). The first row shows \gaia\ $G$-band magnitudes with a limiting magnitude of $G=20.7$ and a minimum proper motion $\mu_{\rm min}$ which depends on magnitude, following the published data release 3 (DR3) dependence \citep[brighter sources can be measured to higher precision,][]{2021A&A...649A...2L}\footnote{\url{https://www.cosmos.esa.int/web/gaia/science-performance}}. The adopted $\mu_{\rm min}$ at each $d$ and A$_{V}$ uses the median magnitude of the companions which are brighter than the limiting magnitude at that $d$ and A$_{V}$. A consequence is that when the number of detectable systems drops to zero, the percentage of the velocity distribution is also zero.

The second and third rows show results for a representative set of \hst\ and \jwst\ observations, with F$160W=26$ and F$277W=26$ limiting magnitudes. To calculate the magnitude dependence of $\mu_{\rm min}$, we construct a toy model as follows. From a 3$\sigma$ limiting magnitude and a target magnitude, we can calculate the signal-to-noise ratio (S/N) of the target. This can be used to estimate the positional uncertainty through $\sigma_{\rm pos}=$ FWHM/(2.35$\times$S/N) \citep{2006obas.book.....B} where FWHM is the full-width at half-maximum of the point spread function. For \hst/WFC3 F160W and \jwst/NIRCam F277W, the FWHM is 0.14\,arcsec and 0.09\,arcsec respectively. The positional uncertainty is added in quadrature to the absolute astrometric uncertainty $\sigma_{\rm abs}$, which for HST / Gaia alignments is typically around 0.5\,mas (for bright stars) if the proper motions of \gaia\ sources are corrected for \citep[i.e., if their positions are moved to match the epoch of the deeper imaging][]{2022ApJ...933...76D,2022arXiv221203256G}. This method can match or even outperform results from differential astrometry \citep{2022ApJ...926..121L}. We adopt floor values for $\sigma_{\rm pos}$ of 0.03 times the pixel scale, a limitation arising from the fact the image is by nature pixelated \citep{2000PASP..112.1383D}. Both the positional and absolute astrometric uncertainties are reduced by a further factor of $\sqrt{N}$, where $N=3$ in these examples, to represent the improvement in source centroiding that arises from image stacking \citep{2022ApJ...933...76D}. This total positional uncertainty $\sigma_{\rm tot}$ is for a single source in one image. The uncertainty on the difference between positions in two images is $\sqrt{2}\sigma_{\rm tot}$ \citep{2022ApJ...933...76D}. If expressed in milliarcseconds, $\mu_{\rm min}$ as a function of magnitude is simply $\sqrt{2}\sigma_{\rm tot}$/N$_{\rm years}$\,mas\,yr$^{-1}$. In Figure \ref{fig:gaiaconstrain} we assume a 5 year baseline between two images.  For \hst\ we use a pixel scale of 0.065\,arcsec\,pixel$^{-1}$ (typical for 3-point dithers after drizzling), and 0.063\,arcsec\,pixel$^{-1}$ for \jwst/NIRCam long-wavelength imaging. Since the minimum measurable proper motion varies linearly with the baseline between images, there is scope for improving these values \citep{2022arXiv221203256G}. We do not account for the decreased performance that will occur for bright objects due to saturation, but note that this can  be mitigated either by shorter exposures, or by moving to \gaia\ (or other observatories) in the regime where this becomes an issue. 

The next row shows results for {\em Euclid}, an upcoming wide-field near-infrared (NIR) observatory with a $\sim$0.7$\times$0.7\,degree field of view. We use the toy model above, a FWHM of 0.3\,arcsec \citep{2011arXiv1110.3193L}, a 5 year baseline and an factor of 10 improvement in positional uncertainty resulting from multiple exposures \citep[$N=100$, where the improvement goes as $\sqrt{N}$,][]{2019JATIS...5d4005W}. The standard Near Infrared Spectrometer and Photometer (NISP) reference limit of $J=24$ is assumed. The final row shows results for the future Nancy Grace Roman Space Telescope (\ngrst), whose Wide Field Imager (WFI) has a field of view 100 times greater than that of \hst\, and the ability to measure proper motions as small as 10\,$\mu$as\,yr$^{-1}$ down to $J=26.7$ \citep{2019JATIS...5d4005W}. We again use the model above, with a FWHM of 0.11\,arcsec, a 5 year baseline and a factor of 10 additional improvement from $N=100$ exposures. Proper motions of 0.01\,mas\,yr$^{-1}$ correspond to transverse velocities as small as 1\,km\,s$^{-1}$ at 20\,kpc, covering nearly the entire velocity distribution of unbound companions, even before considering the magnitude dependence of $\mu_{\rm min}$. {\em Euclid} and \ngrst\ have the added benefit being wide-field - this is a key limitation for \hst\ and \jwst\, which we will return to in the discussion. 

The final column of Figure \ref{fig:gaiaconstrain} shows the product of the first two columns, i.e. the probability at each $d$ and A$_{V}$ that an unbound companion (if present) could be detected {\it and} have its motion measured. We have developed a code using the methodology described in this section to determine this probability, based on the filter, limiting magnitude, distance, extinction and minimum measurable proper motion prescription\footnote{Available at \url{https://github.com/achrimes2/Runaway_probabilities}}. In addition, the probability of any given SNR or young neutron star having an ejected companion must be considered, which we estimate to be $\sim$45 per cent (see Section \ref{sec:explode}). Hence, the probability that a companion will be discovered for any given remnant is given the probability in the third column of Figure \ref{fig:gaiaconstrain} (the fraction of parameter space accessible) multiplied by 0.45 (the probability that the remnant formation ejected a companion). We now apply this code to two example populations: a sample of 13 local SNRs, and Galactic magnetars.

\subsection{Example: local supernova remnants}\label{sec:snrs}
For an example application of the predictions shown in this paper, we refer to a sample of 13 core-collapse supernova remnants studied by \citet{2017A&A...606A..14B} and \citet{2019ApJ...871...92F} \citep[see also][]{2012AdSpR..49.1313F,2014BASI...42...47G,2015MNRAS.448.3196D,2018MNRAS.473.1633K,2019AA...623A..34K,2021AN....342..553L}. They used \gaia\ DR2 to search for stars whose past trajectories intersected the locations of 13 SNRs, and developed a Bayesian framework, based on the expected properties of runaways, to determine the likelihood that each candidate (if any were found) is genuinely associated (rather than a chance crossing). Four of the 13 were found to have a good runaway candidate. Their velocities are shown on Figures \ref{fig:unbounds} and \ref{fig:magvel}, their magnitudes are shown on the relative contribution panel of the \gaia\ row of Figure \ref{fig:photometry_gaiajwst}, and on Figure \ref{fig:magvel}.

Three candidates, first identified by \cite{2017A&A...606A..14B}, are associated with SNRs\,G074.0$-$08.5, G089.0$+$04.7 and G205.5$+$00.5. The stars in question are TYC\,2688-1556$-$1, BD$+$50\,3188 and HD\,261393 respectively. Extinctions, velocities and distances for these stars are drawn from \citet{2017A&A...606A..14B}, their apparent G, G$_{\rm BP}$ and G$_{\rm RP}$ magnitudes are obtained from the \gaia\ DR3 archive. For the fourth runaway candidate, associated with SNR\,G180.0-01.7 (HD\,37424), distance, velocity and extinction information is taken from \citet{2015MNRAS.448.3196D}, and \gaia\ photometry again from the archive. The reported uncertainties account for the photometric, distance and extinction uncertainties. These 13 SNRs are also placed on the five panels of Figure \ref{fig:gaiaconstrain}. Being local SNRs, they cluster in the low $d$, low $A_{\rm V}$ corner of parameter space. We predict that for \gaia\ DR3 with $G<20.7$, 98 per cent of runaway/walkaway parameter is accessible for this SNR sample. Assuming 45 per cent of natal neutron stars have an unbound companion, we therefore predict $0.45 \times 0.98 \times 13 \sim 6$ SNRs should have a discoverable ejected companion. 

For the other four rows in Figure \ref{fig:gaiaconstrain}, the numbers of SNRs for which a runway is expected to be detectable ({\it and} its motion measurable) is also $\sim$6. The lack of improvement from \hst, \jwst, {\em Euclid} and \ngrst\ over \gaia\ is because the sample is sufficiently local that the majority of runaways should already be detectable with \gaia. The prediction of 6 unbound candidates - compared with 4 discovered \citep{2017A&A...606A..14B,2019ApJ...871...92F} - is consistent within Poisson uncertainties. The velocities and magnitudes of the 4 SNR runaway candidates are also in qualitative agreement with our predictions, however, the probability of randomly selecting 4 stars at CMD locations (with non-zero contributions) from the distributions shown in Figure \ref{fig:photometry_gaiajwst} is only 3 per cent (from 10,000 random draws of 4 stars). 

It may be that distance or extinction uncertainties have been underestimated, or even that we are seeing a suggestion of the runaways being slightly redder and more luminous than expected. Firstly, this could arise from the impact of primary supernova ejecta on the secondary. Secondly, it is expected that secondaries post-accretion will typically be redder due to high rotational velocities and equatorial expansion, and over-luminous as they radiate energy to return to thermal equilibrium \citep{1993ASPC...35..207B,1996A&A...305..825V,2021ApJ...923..277R}, but these effects are not yet accounted for in the BPASS models. Therefore, the observation of runaways redder and more luminous than predicted could be due to physics not yet accounted for in these models, but we are prevented from making detailed comparisons due to low number statistics. The scope for building a statistically significant sample of runaways is explored in Section \ref{sec:discuss}.

\begin{figure}
	\includegraphics[width=0.99\columnwidth]{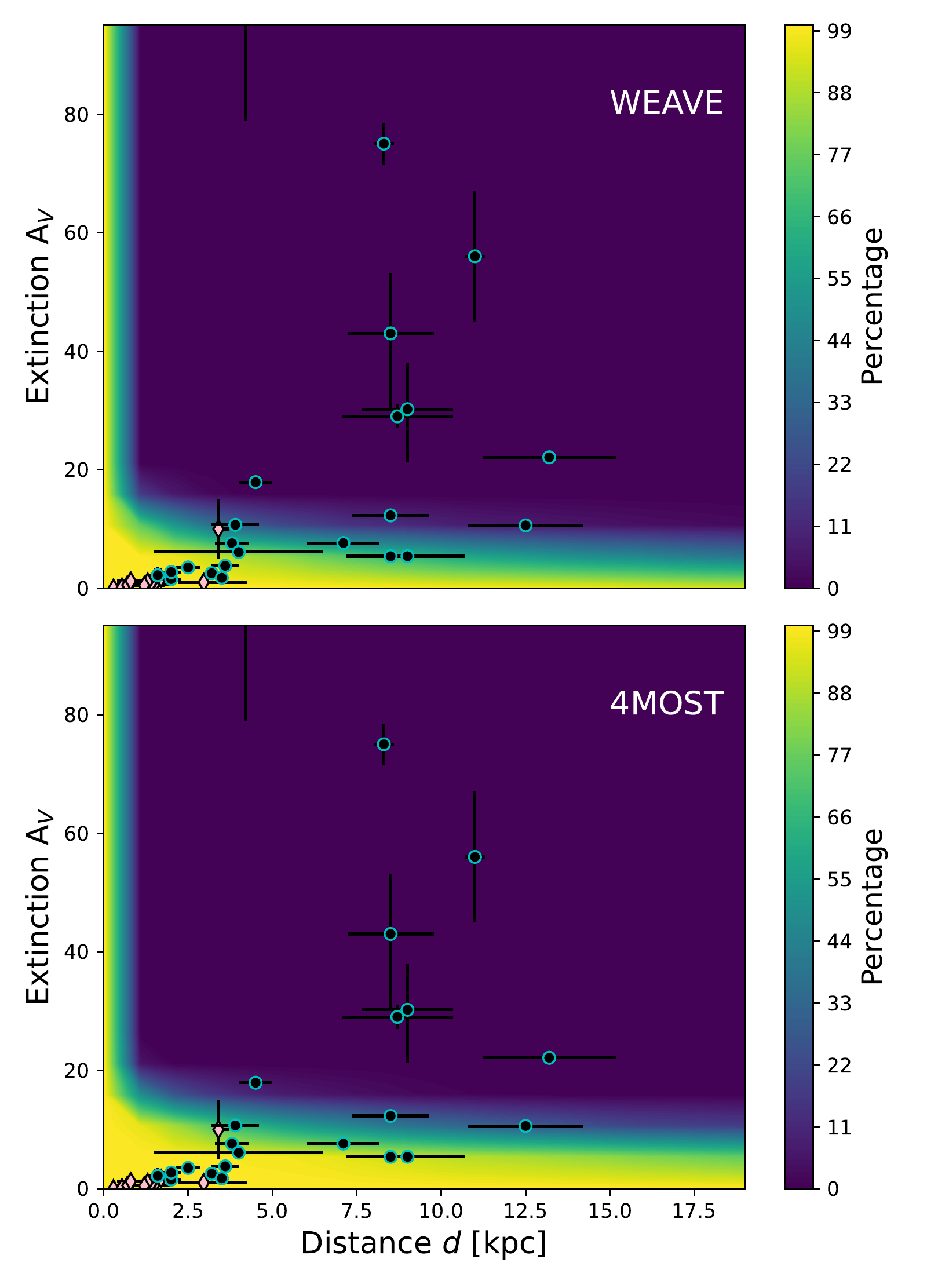}
   \caption{The percentage of the unbound secondary population which is bright enough to be detected at a S/N of 5 or greater in typical WEAVE (top) and 4MOST (bottom) MOS observations (details given in Section \ref{sec:radvel}). In the region of parameter space accessible to \gaia, there are good prospects for radial velocity measurements of candidate runaway stars. SNRs and magnetars are also shown, as in Figure \ref{fig:gaiaconstrain}.}
    \label{fig:radvels}
\end{figure}

\begin{figure*}
	\includegraphics[width=0.49\textwidth]{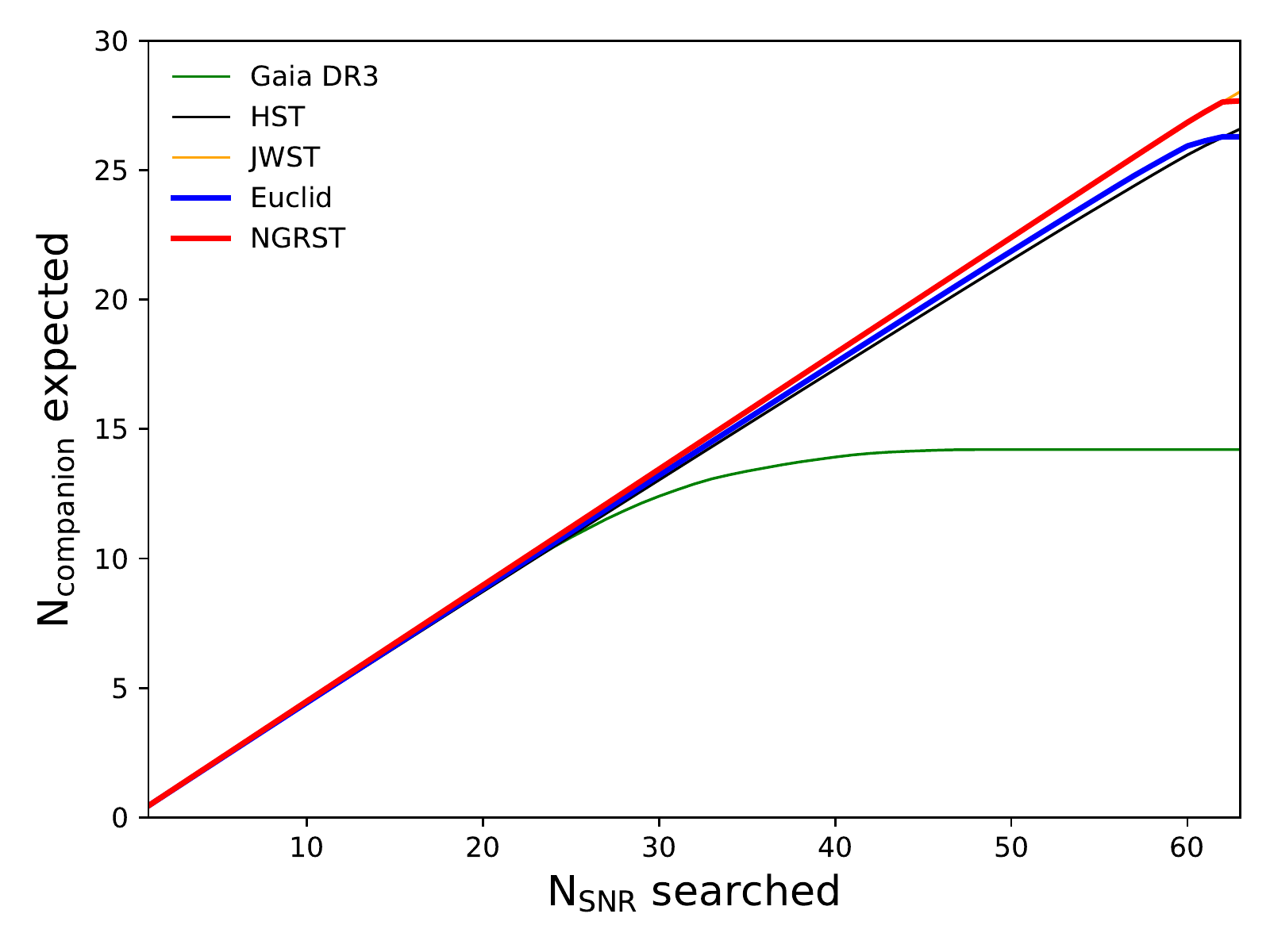}
	\includegraphics[width=0.49\textwidth]{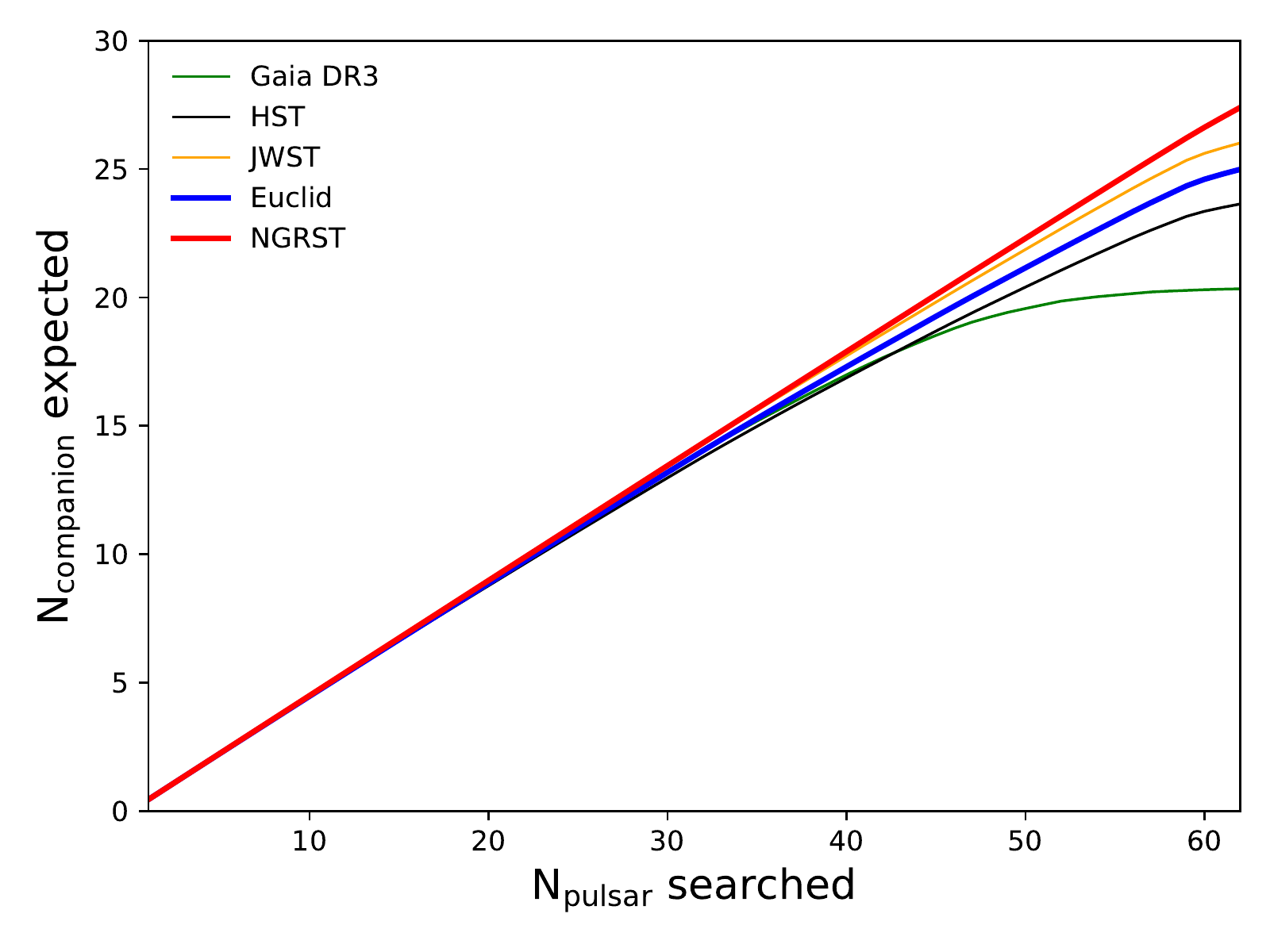}
   \caption{Left: 63 nearby SNRs in the local disc ($d < 5$\,kpc), with distance and extinction estimates from \citet{2020AA...639A..72W}. Using the fiducial model with \citet{2005MNRAS.360..974H} kicks, we calculate the probability for each that a runaway is discoverable for each of the five example scenarios in Figure \ref{fig:gaiaconstrain}. We then rank these from highest to lowest probability, and plot the cumulative distribution. \gaia\ is limited by the image depth, which becomes the limiting factor after the first $\sim$30 objects have been searched, and distances/extinctions become larger. Right: the equivalent figure for Galactic pulsars. Only pulsars with reliable extinction estimates and spin down ages $<10^{5}$\,yr are plotted (N$=62$). We calculate extinctions using the Bayestar 3D dustmaps of \citet{2019ApJ...887...93G}, with dispersion measure distances \citep{1993ApJ...411..674T}  and coordinates from the ANTF pulsar catalogue \citep{2005AJ....129.1993M}.}
    \label{fig:63SNRs}
\end{figure*}

\subsection{Example: Galactic magnetars}
A more challenging neutron star population are the Galactic magnetars, which lie at much greater distances and on dustier sight-lines than the 13 SNRs described above \citep[see][and references therein]{2014ApJS..212....6O}. The estimated distances and extinctions for 23 magnetars \citep[distances and extinctions as listed by][]{2022MNRAS.513.3550C} are plotted on Figure \ref{fig:gaiaconstrain}. For \gaia, we predict $0.24 \times 23 \sim 5$ detections with proper motion. For the \hst, \jwst\ and {\em Euclid} example observations, 8 magnetars are predicted to have discoverable companions. For \ngrst\ the number rises to 9, close to the maximum possible value assuming 45 per cent of supernova eject a non-degenerate companion. For this population, the longer wavelengths and deeper imaging provided by NIR space-based observatories are required to maximise the discovery space. Precise magnetar proper motion measurements are now being made, for example in the radio \citep[][]{2007ApJ...662.1198H} or with deep, high-resolution near-infrared imaging \citep{2012ApJ...761...76T,2013ApJ...772...31T}, revealing that they are currently indistinguishable from pulsars, based on the relatively low numbers of magnetars with good constraints \citet{2022ApJ...926..121L}. A challenge is that the magnetar age range extends up to $\sim$10$^{5}$\,yr (older than SNRs). Propagating over such timescales means that precise proper motions measurements for more magnetars is essential if the past trajectory/birth site are to be determined with sufficient accuracy to identify a small number of possibly associated objects.

\subsection{Radial velocity prospects}\label{sec:radvel}
The previous subsections only considered the prospects for finding unbound secondaries through proper motion surveys, but the measurement of radial velocities (and, if signal-to-noise permits, rotation and chemical abundances) is another tool in our arsenal. We consider two new multi-object spectrographs, WEAVE \citep[on the William Herschel Telescope][]{2014SPIE.9147E..0LD} and 4MOST \citep[on VISTA,][]{2014SPIE.9147E..0MD}.

For WEAVE, we assume a typical observation in high-resolution mode, consisting of 2$\times$1\,hour observing blocks (6 exposures total), and standard observing conditions with a sky brightness in $V$ of 18.5\,mag\,arcsec$^{-2}$ and seeing FWHM of 0.75\,arcsec \footnote{\url{https://www.ing.iac.es/astronomy/instruments/weave/weaveinst}}. Based on the table provided, extrapolating the magnitude versus S/N relation by fitting an exponential gives a S/N$=5$ threshold in the $r$-band of $r=18.9$. We adopt this magnitude as a representative threshold for being able to measure a radial velocity with the WEAVE MOS, and hence show the fraction of unbound secondaries for which WEAVE can measure the radial velocity in Figure \ref{fig:radvels} as a function of distance and extinction. 

For 4MOST, our assumptions are (i) use of the low-resolution spectrographs, (ii) 6$\times$20 minute exposures and (iii) the mean seeing at La Silla \footnote{\url{https://www.4most.eu/cms/facility/capabilities/}} with a new Moon. The limiting AB magnitude for S/N$=5$ at 600nm in these conditions is $\sim$21.2. The percentage of measurable radial velocities (again assuming this requires S/N$>5$ in $r$) is shown in the lower panel of Figure \ref{fig:radvels}. We can see that spectroscopic follow-up of large numbers of candidates in this way is only possible at distances and extinctions also accessible to \gaia, for fainter objects, targeted spectroscopy of individual objects will be required.

\section{Implications \& Discussion}\label{sec:discuss}
\subsection{Prospects for increasing the runaway sample}
We are now in a position to ask how many Galactic runaways and walkaways are discoverable in principle, given the capabilities of current surveys and telescopes. In Figure \ref{fig:63SNRs} we calculate this number for each of the five scenarios shown in Figure \ref{fig:gaiaconstrain}. For SNRs we use the sample of \citet{2020AA...639A..72W}, and their estimates for distances and extinctions. For pulsars we use the ANTF catalogue \citep{2005AJ....129.1993M}\footnote{\url{https://www.atnf.csiro.au/research/pulsar/psrcat}}, adopting dispersion-measure based distances \citep{1993ApJ...411..674T} and line-of-sight extinctions from the Bayestar 3D dust maps \citep{2019ApJ...887...93G}. Only pulsars with reliable extinction estimates and spin-down ages less than $10^{5}$\,yr are shown, since proper motion uncertainties are inflated the longer back a trajectory is traced. For (pessimistic) mas\,yr$^{-1}$ level uncertainties, $10^{5}$\,yr already corresponds to a 100 arcsecond uncertainty in the initial position.

For each of the five rows in Figure \ref{fig:gaiaconstrain}, we calculate the cumulative distribution of the number of runaways discoverable, and find that 14 runaways are in principle traceable back to an associated SNR with \gaia, versus 26 for {\em Euclid}, 27 for \hst\ and 28 for \jwst\ and \ngrst\ (out of 63 SNRs). For pulsars younger than $10^{5}$\,yr it is 20, 24, 25, 26 and 27 (of 62 objects) for \gaia, \hst, {\em Euclid}, \jwst\ and \ngrst\ respectively. In each case, {\em Euclid}, \hst, \jwst\ and \ngrst\ discover almost every unbound companion there to be discovered. For \gaia, no significant improvement is made beyond the first $\sim$30 SNRs or pulsars, where extinction starts to play a significant role in obscuring companions stars in the $G$-band. 

For \hst\ and \jwst, proper motion is the limiting factor, but this doesn't start to become an issue until nearly all SNRs and pulsars in these samples have been searched (i.e. only at the highest distances, $\sim$5\,kpc or greater). Perhaps surprisingly, \jwst\ does not offer a significant improvement over \hst. This is because the benefit of lower extinction at these longer wavelengths only becomes useful at very large extinctions of $A_{\rm V}$>40. {\em Euclid} and \ngrst\ compare well, however {\em Euclid} in particular requires a large number of exposures to reduce positional uncertainties. The primary benefit of {\em Euclid} and \ngrst\ is their field of view, which we will return in this discussion. 

While these observatories open new opportunities, we note that when the SNR and pulsar samples are combined, existing \gaia\ data releases alone offer the possibility of increasing the sample of walkaways/runaways by a factor of $\sim$8. There is however an uncertainty of $\sim$50 per cent on this number, given the discrepancy with (i) other population synthesis codes and (ii) the OB-star runaway fraction \citep[40 per cent predicted here, versus the observed $\sim$20 per cent, e.g.][]{2005A&A...437..247D,2018A&A...616A.149M}.

In the above, we do no restrict the sample to only pulsars with proper motions. To carry out these searches in full requires either a detection of pulsar proper motion or at least an upper limit (also the case for the magnetar sample). It also requires a method to estimate the chance that an association is real (and not a chance crossing); a framework for this has already been developed \citep{2017A&A...606A..14B}. This particularly important since the majority of unbound companions are walkaways, not runaways, and move with velocities comparable to the dispersion of field stars in the Galactic disc \citep{2015A&A...583A..91G}.

\subsection{Spectroscopic follow-up}
A key tool at our disposal to confirm and characterise candidates once identified is spectroscopy. Crucially, it fill in the missing radial velocity component, allowing us to compare the total 3D velocity distributions as predicted by population synthesis with observations. The prospects for obtaining radial velocities for stars with proper motion constraints were explored in Section \ref{sec:radvel} and Figure \ref{fig:radvels}. Spectroscopy can also be used to search for high rotational velocities, which might be preferentially expected among ejected companions (compared with dynamically ejected stars) due to mass transfer \citep{1991A&A...241..419P,1993ASPC...35..207B,2018MNRAS.477.5261B,2019A&A...624A..66R,2022arXiv221113476S}. Furthermore, past interactions can lead to unusual chemical compositions which may make ejected companions stand out with respect to field stars \citep[e.g.][]{2014A&A...565A..90C,2021ApJ...923..277R,2021AJ....161..248W}. Multi-object spectrographs (MOS) and Integral Field Unit (IFU) instruments, notably the upcoming wide-filed WEAVE and 4MOST, will be able to characterise numerous candidates identified from proper motion surveys, providing key additional information. A caveat for these ground-based observations is the limiting magnitude. For WEAVE's high-resolution and low-resolution MOS modes, the faintest accessible objects in a 1 hour exposure have $V\sim18$ and $\sim21$ respectively \citep{2014SPIE.9147E..0LD}. This is comparable to the limiting magnitude of \gaia\ (see the upper left panel of Figure \ref{fig:gaiaconstrain}), so follow-up of \gaia\ candidates will be possible, but challenging for fainter sources. 4MOST can also obtain obtain radial velocities for objects down to the \gaia\ limiting magnitude, and stellar parameters for objects brighter than 18$^{\rm th}$ magnitude \citep[with 2 hours of exposure time, and depending on sky conditions,][]{2019Msngr.175....3D}. For MOS and IFU spectroscopy of the faintest targets, \jwst/NIRSpec can reach 23-24$^{\rm th}$ magnitude (in 10\,ks) across the 1--4$\mu$m range \citep{2022SPIE12180E..0XG}.

\subsection{Field of view considerations}
A practical limitation of using observatories such as \jwst\ and \hst\ is the field of view (FOV), which is approximately 2.2 and 2.0 arcmin for HST/WFC3 and JWST/NIRCam respectively. For the following we assume a transverse velocity of 100\,km\,s$^{-1}$, which covers $\sim$90 per cent of the distribution (see Figure \ref{fig:unbounds}). For remnants of age 10$^{5}$\,yr, the angular offset of an unbound companion travelling at 100\,km\,s$^{-1}$ in the plane of the sky is greater than half the FOV of \hst\ and \jwst\ as far out as 20\,kpc. We use half the FOV here as a guide, assuming that the SNR or neutron star position at birth is centred in the FOV. Therefore, some degree of image tiling would be necessary - a three by three grid would be required even at $\sim$10\,kpc. For \ngrst, however, a single FOV is sufficient for distances beyond $\sim$2.5\,kpc - and for distances less than this, \gaia\ is typically capable of discovering most unbound companions. If the remnant is a factor of 10 younger, then the 100\,km\,s$^{-1}$ runaway will also be within half a \hst\ or \jwst\ FOV beyond $\sim$2.5\,kpc. Hence, while \hst\ and \jwst\ do have a field of view constraint, they can be applied for unbound companion searches, without tiling, if the remnant is young and distant (and hence small on the plane of the sky). For nearby objects, \gaia\ remains the best tool at our disposal, while for older remnants beyond the local few kpc, wide-field NIR observatories such as \ngrst\ or {\em Euclid} \citep{2011arXiv1110.3193L} will be required. Although these missions are cosmology-focused, and will preferentially avoid observing the Galactic plane, their application to runaway searches \citep[and other Galactic science,][]{2019JATIS...5d4005W} may be possible through general observing time programmes. A \gaia-like observatory operating in the NIR \citep{2016arXiv160907325H} would also greatly increase the parameter space in which unbound companions could be discovered.

\subsection{Modelling approximations}
\citet{2011MNRAS.414.3501E} found that a few per cent of OB-stars have runaway velocities ($>30$\,km\,s$^{-1}$), well below observational estimates. The updated models of \citet{2017PASA...34...58E} and \citet{2017ApJS..230...15M} binary parameter distributions used in this work instead produce an OB runaway fraction of $\sim$40 per cent \citep[compared observational estimates of around 10--20 per cent, e.g.][]{2018A&A...616A.149M}. As noted in Section \ref{sec:unbounds}, the high rate of RLOF to CEE progression in BPASS v2 - due to a combination of the input parameter distributions and how mass transfer is handled - produces a large number of tight binaries at core-collapse, which is driving the high velocities predicted.  

The choice of binary population synthesis input binary parameters naturally has an impact on the velocity distribution outputs. The impact of (statistical) binary parameter uncertainty in BPASS was explored in the context of spectral synthesis by \citet{2020MNRAS.495.4605S}, and in the context of bound companion predictions by \citet[][where uncertainties in the input parameters lead to uncertainties in the bound fraction of a few per cent]{2022MNRAS.513.3550C}. \citet{2019A&A...624A..66R} trial a range of binary parameter distributions, finding that (in terms of the ratio of walkways to runaways for example) the specific input distribution choices can change the results by as much as $\sim$50 per cent. A full investigation of binary parameter uncertainty on the results presented here is beyond the scope of this paper, but we note that {\it within} codes there is significant uncertainty, and even more {\it between} them, where additional differences primarily arise due to different prescriptions for mass loss rates, mass transfer efficiency, stability and the treatment of the common envelope phase. Overall, there is a discrepancy of up to a factor of few in velocity between population synthesis codes and observations (at least for OB stars), and it is in this context that our predictions should be interpreted. 

Throughout, we have assumed a \gaia\ limiting magnitude of $G=20.7$. However, we note that the survey completeness at fixed limiting magnitude is a function (primarily) of Galactic longitude and latitude. This arises due to a complex combination of the \gaia\ scanning law and a lower efficiency of faint source detections in crowded regions. This is discussed in detail by \citet{2022arXiv220809335C}, \citet{2020MNRAS.497.4246B} and references therein.

Our examples and projections for the numbers of runaways discoverable assume that the secondary has not gone supernova before the time of observation. For young Galactic objects, such as SNRs, magnetars and young pulsars less than $10^{5}$\,yr old, the probability of the secondary going supernova before the present epoch is extremely low, with the shortest-lived post-supernova secondary BPASS model lasting 3\,Myr after disruption. This is accounting for the occurrence of rejuvenation in some systems. Rejuvenation requires enhanced mixing, which can occur in a variety of ways \citep[e.g. mass transfer increasing the temperature gradient and driving stronger core convection in the accretor][]{2023ApJ...942L..32R}. In BPASS, rejuvenation is assumed to occur due to spin-up and rotational mixing \citep[see][for a full discussion]{2023MNRAS.518..860G} and is triggered when the secondary accretes more than 5 per cent of its initial mass and has M$>$2\,M$_{\odot}$. At this point, the secondary model is replaced with ZAMS model with the new initial mass \citep{2019MNRAS.482..870E}.

Another uncertainty is in the progenitor systems and whether the kick velocity is dependent on the progenitor. For instance, ECSNe may produce lower velocity kicks \citep[e.g.][]{2006ApJ...644.1063D,2012ARNPS..62..407J,2018MNRAS.480.2011G}, but here we have randomly assigned kicks independent of the progenitor properties or the type of supernova. Throughout, we have solely focused on supernovae producing neutron stars, but if successful supernovae can also produce black holes, then a full analysis should include their formation too. This may be particularly relevant for SNe Ibc, since self-stripped single stars are unlikely to produce successful supernovae \citep{2021A&A...656L..19Z}. Given the complex landscape of explodability \citep[e.g.][]{2020MNRAS.499.2803P} and the uncertain kick distribution of natal black holes, we leave such analyses to future work. Although around $\sim$20 per cent of core-collapse events are expected to produce a black hole, the fraction of successful supernovae expected to produce black holes (through fallback onto a natal neutron star) is expected to be rare \citep[e.g.][]{2015MNRAS.446.1213K,2020ApJ...890...51E}

\subsection{Future observational tests}
In principle, if supernova runaway sample sizes grow large enough, many new tests of binary evolution and supernovae will become possible. For instance, in principle it should be possible to compare the composition of SNRs with the properties of their associated runaways. We can ask, for example, whether SNRs from stripped envelope events produce faster and more massive runaways (see Figure \ref{fig:magvel}. While it is possible to distinguish core-collapse from thermonuclear SNRs, separating type II and type Ibc core-collapse remnants is challenging. Constraints may be obtainable through SNR oxygen abundances as a progenitor mass proxy \citep{2012A&ARv..20...49V}, but this picture is complicated by binary interactions which can ultimately alter the core composition \citep{2020A&A...637A...6L,2021A&A...656A..58L}.

Finally, we examine the impact of using different kick distributions on the expected results from the 13 SNRs as described in Section \ref{sec:snrs}. Switching to Verbunt kicks and propagating through, we predict that 4.7 of the 13 SNRs of \citet{2017A&A...606A..14B} and \citet{2019ApJ...871...92F} would have detectable companions, moving fast enough for their motion to be measured, versus 4.8 when the \citet{2005MNRAS.360..974H} kick distribution is assumed. Using the Bray kick, the predicted number for the 13 SNRs searched with \gaia\ is 3.7. In the current realm of small number statistics, it is difficult to obtain constraints on the kick in this way, as all three estimates are consistent (within Poisson errors) with 4 candidates being discovered. However, if large samples of up to $\sim$50 can be searched (Fig. \ref{fig:63SNRs}), then the differences will be pronounced. The distribution of angles between proper motion vectors also contains useful information, where stronger peaks indicate stronger natal kicks. In summary, there is potential for using the kinematics of unbound companions as an independent constraint on neutron natal kicks and the dynamics of supernovae in binaries.

\section{Conclusions}
In this work, we have produced kinematic and photometric predictions for stars ejected by the supernova of their binary companions. We find a runaway fraction among ejected OB stars of 40 per cent. These predictions are placed in the context of existing unbound companion searches and predictions from other population synthesis codes, in addition to current and upcoming observational opportunities. We have shown that the velocities of neutron stars produced in a binaries closely follow the input kick distribution, justifying the assumption that observed neutron star velocities are directly representative of their natal kicks. We have demonstrated that for optical surveys such as \gaia, the limiting factor in discovering unbound companions is the survey depth. However, we estimate that the numbers can still be increased by a factor of 5--10 with existing \gaia\ data releases. The discovery space can be increased further if observatories capable of deep IR imaging such as \hst\ and \jwst\ are employed, for which the primary limiting factor is the instrumental field of view. Discoveries can be maximised with upcoming wide-field NIR observatories such as {\em Euclid} and \ngrst. Candidates can then be differentiated from unassociated field stars through the identification of high-velocities, rapidly rotation and chemically anomalous compositions with multi-object spectrographs and integral field units such as WEAVE and 4MOST. These have wide fields of view and high enough spectral resolution to sample nearly the entire runaway and walkaway velocity distribution within the Galaxy. If the number of ejected companion candidates is increased, as we have shown is feasible with current and upcoming facilities, new approaches to constrain the binary progenitors of supernovae, the dynamics of supernovae in binaries and neutron star natal kicks will become possible.

\section*{Acknowledgements}
This work is part of the research programme Athena with project number 184.034.002, which is (partly) financed by the Dutch Research Council (NWO). MF is supported by a Royal Society - Science Foundation Ireland University Research Fellowship.

This work made use of v2.2.1 of the Binary Population and Spectral Synthesis (BPASS) models as described in \citet{2017PASA...34...58E} and \citet{2018MNRAS.479...75S}, and the python package for BPASS, HOKI \citep{2020JOSS....5.1987S}. This work has made use of {\sc ipython} \citep{2007CSE.....9c..21P}, {\sc numpy} \citep{2020arXiv200610256H}, {\sc scipy} \citep{2020NatMe..17..261V}; {\sc matplotlib} \citep{2007CSE.....9...90H}, Seaborn packages \citep{Waskom2021} and {\sc astropy},\footnote{https://www.astropy.org} a community-developed core Python package for Astronomy \citep{astropy:2013, astropy:2018}. We have also made use of the python modules {\sc dustmaps} \citep{2018JOSS....3..695M} and {\sc extinction} \citep{barbary_kyle_2016_804967}. This research has made use of the Spanish Virtual Observatory (\url{https://svo.cab.inta-csic.es}) project funded by MCIN/AEI/10.13039/501100011033/ through grant PID2020-112949GB-I00 \citep{2012ivoa.rept.1015R,2020sea..confE.182R}. This work has made use of data from the European Space Agency (ESA) mission {\it Gaia} (\url{https://www.cosmos.esa.int/gaia}), processed by the {\it Gaia} Data Processing and Analysis Consortium (DPAC, \url{https://www.cosmos.esa.int/web/gaia/dpac/consortium}). Funding for the DPAC has been provided by national institutions, in particular the institutions participating in the {\it Gaia} Multilateral Agreement.

Finally, we thank the referee for their constructive comments on this manuscript.

\section*{Data Availability}
The code used to calculate the probability of finding runaways from Galactic objects is available at \url{https://github.com/achrimes2/Runaway_probabilities}. The BPASS binary evolution models used in this work are available via the project website: \url{https://bpass.auckland.ac.nz/}.



\bibliographystyle{mnras}
\bibliography{paper} 

\begin{thebibliography}{}
\makeatletter
\relax
\def\mn@urlcharsother{\let\do\@makeother \do\$\do\&\do\#\do\^\do\_\do\%\do\~}
\def\mn@doi{\begingroup\mn@urlcharsother \@ifnextchar [ {\mn@doi@}
  {\mn@doi@[]}}
\def\mn@doi@[#1]#2{\def\@tempa{#1}\ifx\@tempa\@empty \href
  {http://dx.doi.org/#2} {doi:#2}\else \href {http://dx.doi.org/#2} {#1}\fi
  \endgroup}
\def\mn@eprint#1#2{\mn@eprint@#1:#2::\@nil}
\def\mn@eprint@arXiv#1{\href {http://arxiv.org/abs/#1} {{\tt arXiv:#1}}}
\def\mn@eprint@dblp#1{\href {http://dblp.uni-trier.de/rec/bibtex/#1.xml}
  {dblp:#1}}
\def\mn@eprint@#1:#2:#3:#4\@nil{\def\@tempa {#1}\def\@tempb {#2}\def\@tempc
  {#3}\ifx \@tempc \@empty \let \@tempc \@tempb \let \@tempb \@tempa \fi \ifx
  \@tempb \@empty \def\@tempb {arXiv}\fi \@ifundefined
  {mn@eprint@\@tempb}{\@tempb:\@tempc}{\expandafter \expandafter \csname
  mn@eprint@\@tempb\endcsname \expandafter{\@tempc}}}

\bibitem[\protect\citeauthoryear{{Abt} \& {Levy}}{{Abt} \&
  {Levy}}{1976}]{1976ApJS...30..273A}
{Abt} H.~A.,  {Levy} S.~G.,  1976, \mn@doi [\apjs] {10.1086/190363}, \href
  {https://ui.adsabs.harvard.edu/abs/1976ApJS...30..273A} {30, 273}

\bibitem[\protect\citeauthoryear{{Astropy Collaboration} et~al.,}{{Astropy
  Collaboration} et~al.}{2013}]{astropy:2013}
{Astropy Collaboration} et~al., 2013, \mn@doi [\aap]
  {10.1051/0004-6361/201322068}, \href
  {http://adsabs.harvard.edu/abs/2013A%26A...558A..33A} {558, A33}

\bibitem[\protect\citeauthoryear{Barbary}{Barbary}{2016}]{barbary_kyle_2016_804967}
Barbary K.,  2016, extinction v0.3.0, \mn@doi{10.5281/zenodo.804967}, \url
  {https://doi.org/10.5281/zenodo.804967}

\bibitem[\protect\citeauthoryear{{Belczynski}, {Kalogera}, {Rasio}, {Taam},
  {Zezas}, {Bulik}, {Maccarone}  \& {Ivanova}}{{Belczynski}
  et~al.}{2008}]{2008ApJS..174..223B}
{Belczynski} K.,  {Kalogera} V.,  {Rasio} F.~A.,  {Taam} R.~E.,  {Zezas} A.,
  {Bulik} T.,  {Maccarone} T.~J.,   {Ivanova} N.,  2008, \mn@doi [\apjs]
  {10.1086/521026}, \href
  {https://ui.adsabs.harvard.edu/abs/2008ApJS..174..223B} {174, 223}

\bibitem[\protect\citeauthoryear{{Birney}, {Gonzalez}  \& {Oesper}}{{Birney}
  et~al.}{2006}]{2006obas.book.....B}
{Birney} D.~S.,  {Gonzalez} G.,   {Oesper} D.,  2006, {Observational Astronomy
  - 2nd Edition}.
Cambridge University Press, \mn@doi{10.2277/0521853702}

\bibitem[\protect\citeauthoryear{{Blaauw}}{{Blaauw}}{1961}]{1961BAN....15..265B}
{Blaauw} A.,  1961, \bain, \href
  {https://ui.adsabs.harvard.edu/abs/1961BAN....15..265B} {15, 265}

\bibitem[\protect\citeauthoryear{{Blaauw}}{{Blaauw}}{1993}]{1993ASPC...35..207B}
{Blaauw} A.,  1993, in {Cassinelli} J.~P.,  {Churchwell} E.~B.,  eds,
  Astronomical Society of the Pacific Conference Series Vol. 35, Massive Stars:
  Their Lives in the Interstellar Medium. p.~207

\bibitem[\protect\citeauthoryear{{Boersma}}{{Boersma}}{1961}]{1961BAN....15..291B}
{Boersma} J.,  1961, \bain, \href
  {https://ui.adsabs.harvard.edu/abs/1961BAN....15..291B} {15, 291}

\bibitem[\protect\citeauthoryear{{Boubert} \& {Evans}}{{Boubert} \&
  {Evans}}{2018}]{2018MNRAS.477.5261B}
{Boubert} D.,  {Evans} N.~W.,  2018, \mn@doi [\mnras] {10.1093/mnras/sty980},
  \href {https://ui.adsabs.harvard.edu/abs/2018MNRAS.477.5261B} {477, 5261}

\bibitem[\protect\citeauthoryear{{Boubert} \& {Everall}}{{Boubert} \&
  {Everall}}{2020}]{2020MNRAS.497.4246B}
{Boubert} D.,  {Everall} A.,  2020, \mn@doi [\mnras] {10.1093/mnras/staa2305},
  \href {https://ui.adsabs.harvard.edu/abs/2020MNRAS.497.4246B} {497, 4246}

\bibitem[\protect\citeauthoryear{{Boubert}, {Erkal}, {Evans}  \&
  {Izzard}}{{Boubert} et~al.}{2017a}]{2017MNRAS.469.2151B}
{Boubert} D.,  {Erkal} D.,  {Evans} N.~W.,   {Izzard} R.~G.,  2017a, \mn@doi
  [\mnras] {10.1093/mnras/stx848}, \href
  {https://ui.adsabs.harvard.edu/abs/2017MNRAS.469.2151B} {469, 2151}

\bibitem[\protect\citeauthoryear{{Boubert}, {Fraser}, {Evans}, {Green}  \&
  {Izzard}}{{Boubert} et~al.}{2017b}]{2017A&A...606A..14B}
{Boubert} D.,  {Fraser} M.,  {Evans} N.~W.,  {Green} D.~A.,   {Izzard} R.~G.,
  2017b, \mn@doi [\aap] {10.1051/0004-6361/201731142}, \href
  {https://ui.adsabs.harvard.edu/abs/2017A&A...606A..14B} {606, A14}

\bibitem[\protect\citeauthoryear{{Bray} \& {Eldridge}}{{Bray} \&
  {Eldridge}}{2016}]{2016MNRAS.461.3747B}
{Bray} J.~C.,  {Eldridge} J.~J.,  2016, \mn@doi [\mnras]
  {10.1093/mnras/stw1275}, \href
  {https://ui.adsabs.harvard.edu/abs/2016MNRAS.461.3747B} {461, 3747}

\bibitem[\protect\citeauthoryear{{Bray} \& {Eldridge}}{{Bray} \&
  {Eldridge}}{2018}]{2018MNRAS.480.5657B}
{Bray} J.~C.,  {Eldridge} J.~J.,  2018, \mn@doi [\mnras]
  {10.1093/mnras/sty2230}, \href
  {https://ui.adsabs.harvard.edu/abs/2018MNRAS.480.5657B} {480, 5657}

\bibitem[\protect\citeauthoryear{{Briel}, {Eldridge}, {Stanway}, {Stevance}  \&
  {Chrimes}}{{Briel} et~al.}{2022}]{2022MNRAS.514.1315B}
{Briel} M.~M.,  {Eldridge} J.~J.,  {Stanway} E.~R.,  {Stevance} H.~F.,
  {Chrimes} A.~A.,  2022, \mn@doi [\mnras] {10.1093/mnras/stac1100}, \href
  {https://ui.adsabs.harvard.edu/abs/2022MNRAS.514.1315B} {514, 1315}

\bibitem[\protect\citeauthoryear{{Briel}, {Stevance}  \& {Eldridge}}{{Briel}
  et~al.}{2023}]{2023MNRAS.520.5724B}
{Briel} M.~M.,  {Stevance} H.~F.,   {Eldridge} J.~J.,  2023, \mn@doi [\mnras]
  {10.1093/mnras/stad399}, \href
  {https://ui.adsabs.harvard.edu/abs/2023MNRAS.520.5724B} {520, 5724}

\bibitem[\protect\citeauthoryear{{Cantat-Gaudin} et~al.,}{{Cantat-Gaudin}
  et~al.}{2022}]{2022arXiv220809335C}
{Cantat-Gaudin} T.,  et~al., 2022, arXiv e-prints, \href
  {https://ui.adsabs.harvard.edu/abs/2022arXiv220809335C} {p. arXiv:2208.09335}

\bibitem[\protect\citeauthoryear{{Chrimes}, {Stanway}  \& {Eldridge}}{{Chrimes}
  et~al.}{2020}]{2020MNRAS.491.3479C}
{Chrimes} A.~A.,  {Stanway} E.~R.,   {Eldridge} J.~J.,  2020, \mn@doi [\mnras]
  {10.1093/mnras/stz3246}, \href
  {https://ui.adsabs.harvard.edu/abs/2020MNRAS.491.3479C} {491, 3479}

\bibitem[\protect\citeauthoryear{{Chrimes} et~al.,}{{Chrimes}
  et~al.}{2022}]{2022MNRAS.513.3550C}
{Chrimes} A.~A.,  et~al., 2022, \mn@doi [\mnras] {10.1093/mnras/stac1090},
  \href {https://ui.adsabs.harvard.edu/abs/2022MNRAS.513.3550C} {513, 3550}

\bibitem[\protect\citeauthoryear{{Clark}, {Goodwin}, {Crowther}, {Kaper},
  {Fairbairn}, {Langer}  \& {Brocksopp}}{{Clark}
  et~al.}{2002}]{2002A&A...392..909C}
{Clark} J.~S.,  {Goodwin} S.~P.,  {Crowther} P.~A.,  {Kaper} L.,  {Fairbairn}
  M.,  {Langer} N.,   {Brocksopp} C.,  2002, \mn@doi [\aap]
  {10.1051/0004-6361:20021184}, \href
  {https://ui.adsabs.harvard.edu/abs/2002A&A...392..909C} {392, 909}

\bibitem[\protect\citeauthoryear{{Clark}, {Ritchie}, {Najarro}, {Langer}  \&
  {Negueruela}}{{Clark} et~al.}{2014}]{2014A&A...565A..90C}
{Clark} J.~S.,  {Ritchie} B.~W.,  {Najarro} F.,  {Langer} N.,   {Negueruela}
  I.,  2014, \mn@doi [\aap] {10.1051/0004-6361/201321771}, \href
  {https://ui.adsabs.harvard.edu/abs/2014A&A...565A..90C} {565, A90}

\bibitem[\protect\citeauthoryear{{Dalton} et~al.,}{{Dalton}
  et~al.}{2014}]{2014SPIE.9147E..0LD}
{Dalton} G.,  et~al., 2014, in {Ramsay} S.~K.,  {McLean} I.~S.,   {Takami} H.,
  eds,  Society of Photo-Optical Instrumentation Engineers (SPIE) Conference
  Series Vol. 9147, Ground-based and Airborne Instrumentation for Astronomy V.
  p. 91470L (\mn@eprint {arXiv} {1412.0843}), \mn@doi{10.1117/12.2055132}

\bibitem[\protect\citeauthoryear{{De Donder}, {Vanbeveren}  \& {van Bever}}{{De
  Donder} et~al.}{1997}]{1997A&A...318..812D}
{De Donder} E.,  {Vanbeveren} D.,   {van Bever} J.,  1997, \aap, \href
  {https://ui.adsabs.harvard.edu/abs/1997A&A...318..812D} {318, 812}

\bibitem[\protect\citeauthoryear{{Dessart}, {Burrows}, {Ott}, {Livne}, {Yoon}
  \& {Langer}}{{Dessart} et~al.}{2006}]{2006ApJ...644.1063D}
{Dessart} L.,  {Burrows} A.,  {Ott} C.~D.,  {Livne} E.,  {Yoon} S.~C.,
  {Langer} N.,  2006, \mn@doi [\apj] {10.1086/503626}, \href
  {https://ui.adsabs.harvard.edu/abs/2006ApJ...644.1063D} {644, 1063}

\bibitem[\protect\citeauthoryear{{Dessart}, {Hillier}, {Livne}, {Yoon},
  {Woosley}, {Waldman}  \& {Langer}}{{Dessart}
  et~al.}{2011}]{2011MNRAS.414.2985D}
{Dessart} L.,  {Hillier} D.~J.,  {Livne} E.,  {Yoon} S.-C.,  {Woosley} S.,
  {Waldman} R.,   {Langer} N.,  2011, \mn@doi [\mnras]
  {10.1111/j.1365-2966.2011.18598.x}, \href
  {https://ui.adsabs.harvard.edu/abs/2011MNRAS.414.2985D} {414, 2985}

\bibitem[\protect\citeauthoryear{{Din{\c{c}}el}, {Neuh{\"a}user}, {Yerli},
  {Ankay}, {Tetzlaff}, {Torres}  \& {Mugrauer}}{{Din{\c{c}}el}
  et~al.}{2015}]{2015MNRAS.448.3196D}
{Din{\c{c}}el} B.,  {Neuh{\"a}user} R.,  {Yerli} S.~K.,  {Ankay} A.,
  {Tetzlaff} N.,  {Torres} G.,   {Mugrauer} M.,  2015, \mn@doi [\mnras]
  {10.1093/mnras/stv124}, \href
  {https://ui.adsabs.harvard.edu/abs/2015MNRAS.448.3196D} {448, 3196}

\bibitem[\protect\citeauthoryear{{Dolphin}}{{Dolphin}}{2000}]{2000PASP..112.1383D}
{Dolphin} A.~E.,  2000, \mn@doi [\pasp] {10.1086/316630}, \href
  {https://ui.adsabs.harvard.edu/abs/2000PASP..112.1383D} {112, 1383}

\bibitem[\protect\citeauthoryear{{Eldridge} \& {Tout}}{{Eldridge} \&
  {Tout}}{2004}]{2004MNRAS.353...87E}
{Eldridge} J.~J.,  {Tout} C.~A.,  2004, \mn@doi [\mnras]
  {10.1111/j.1365-2966.2004.08041.x}, \href
  {https://ui.adsabs.harvard.edu/abs/2004MNRAS.353...87E} {353, 87}

\bibitem[\protect\citeauthoryear{{Eldridge}, {Langer}  \& {Tout}}{{Eldridge}
  et~al.}{2011}]{2011MNRAS.414.3501E}
{Eldridge} J.~J.,  {Langer} N.,   {Tout} C.~A.,  2011, \mn@doi [\mnras]
  {10.1111/j.1365-2966.2011.18650.x}, \href
  {https://ui.adsabs.harvard.edu/abs/2011MNRAS.414.3501E} {414, 3501}

\bibitem[\protect\citeauthoryear{{Eldridge}, {Stanway}, {Xiao}, {McClelland},
  {Taylor}, {Ng}, {Greis}  \& {Bray}}{{Eldridge}
  et~al.}{2017}]{2017PASA...34...58E}
{Eldridge} J.~J.,  {Stanway} E.~R.,  {Xiao} L.,  {McClelland} L.~A.~S.,
  {Taylor} G.,  {Ng} M.,  {Greis} S.~M.~L.,   {Bray} J.~C.,  2017, \mn@doi
  [\pasa] {10.1017/pasa.2017.51}, \href
  {https://ui.adsabs.harvard.edu/abs/2017PASA...34...58E} {34, e058}

\bibitem[\protect\citeauthoryear{{Eldridge}, {Stanway}  \& {Tang}}{{Eldridge}
  et~al.}{2019}]{2019MNRAS.482..870E}
{Eldridge} J.~J.,  {Stanway} E.~R.,   {Tang} P.~N.,  2019, \mn@doi [\mnras]
  {10.1093/mnras/sty2714}, \href
  {https://ui.adsabs.harvard.edu/abs/2019MNRAS.482..870E} {482, 870}

\bibitem[\protect\citeauthoryear{{Ertl}, {Woosley}, {Sukhbold}  \&
  {Janka}}{{Ertl} et~al.}{2020}]{2020ApJ...890...51E}
{Ertl} T.,  {Woosley} S.~E.,  {Sukhbold} T.,   {Janka} H.~T.,  2020, \mn@doi
  [\apj] {10.3847/1538-4357/ab6458}, \href
  {https://ui.adsabs.harvard.edu/abs/2020ApJ...890...51E} {890, 51}

\bibitem[\protect\citeauthoryear{{Evans}, {Renzo}  \& {Rossi}}{{Evans}
  et~al.}{2020}]{2020MNRAS.497.5344E}
{Evans} F.~A.,  {Renzo} M.,   {Rossi} E.~M.,  2020, \mn@doi [\mnras]
  {10.1093/mnras/staa2334}, \href
  {https://ui.adsabs.harvard.edu/abs/2020MNRAS.497.5344E} {497, 5344}

\bibitem[\protect\citeauthoryear{{Fang}, {Maeda}, {Kuncarayakti}, {Sun}  \&
  {Gal-Yam}}{{Fang} et~al.}{2019}]{2019NatAs...3..434F}
{Fang} Q.,  {Maeda} K.,  {Kuncarayakti} H.,  {Sun} F.,   {Gal-Yam} A.,  2019,
  \mn@doi [Nature Astronomy] {10.1038/s41550-019-0710-6}, \href
  {https://ui.adsabs.harvard.edu/abs/2019NatAs...3..434F} {3, 434}

\bibitem[\protect\citeauthoryear{{Ferrand} \& {Safi-Harb}}{{Ferrand} \&
  {Safi-Harb}}{2012}]{2012AdSpR..49.1313F}
{Ferrand} G.,  {Safi-Harb} S.,  2012, \mn@doi [Advances in Space Research]
  {10.1016/j.asr.2012.02.004}, \href
  {https://ui.adsabs.harvard.edu/abs/2012AdSpR..49.1313F} {49, 1313}

\bibitem[\protect\citeauthoryear{{Filippenko}}{{Filippenko}}{1997}]{1997ARA&A..35..309F}
{Filippenko} A.~V.,  1997, \mn@doi [\araa] {10.1146/annurev.astro.35.1.309},
  \href {https://ui.adsabs.harvard.edu/abs/1997ARA&A..35..309F} {35, 309}

\bibitem[\protect\citeauthoryear{{Fortin}, {Garcia}, {Chaty},
  {Chassande-Mottin}  \& {Simaz Bunzel}}{{Fortin}
  et~al.}{2022}]{2022arXiv220603904F}
{Fortin} F.,  {Garcia} F.,  {Chaty} S.,  {Chassande-Mottin} E.,   {Simaz
  Bunzel} A.,  2022, arXiv e-prints, \href
  {https://ui.adsabs.harvard.edu/abs/2022arXiv220603904F} {p. arXiv:2206.03904}

\bibitem[\protect\citeauthoryear{{Fraser} \& {Boubert}}{{Fraser} \&
  {Boubert}}{2019}]{2019ApJ...871...92F}
{Fraser} M.,  {Boubert} D.,  2019, \mn@doi [\apj] {10.3847/1538-4357/aaf6b8},
  \href {https://ui.adsabs.harvard.edu/abs/2019ApJ...871...92F} {871, 92}

\bibitem[\protect\citeauthoryear{{Freire}, {Ransom}, {B{\'e}gin}, {Stairs},
  {Hessels}, {Frey}  \& {Camilo}}{{Freire} et~al.}{2008}]{2008ApJ...675..670F}
{Freire} P. C.~C.,  {Ransom} S.~M.,  {B{\'e}gin} S.,  {Stairs} I.~H.,
  {Hessels} J. W.~T.,  {Frey} L.~H.,   {Camilo} F.,  2008, \mn@doi [\apj]
  {10.1086/526338}, \href
  {https://ui.adsabs.harvard.edu/abs/2008ApJ...675..670F} {675, 670}

\bibitem[\protect\citeauthoryear{{Gaia Collaboration} et~al.,}{{Gaia
  Collaboration} et~al.}{2021}]{2021A&A...649A...6G}
{Gaia Collaboration} et~al., 2021, \mn@doi [\aap]
  {10.1051/0004-6361/202039498}, \href
  {https://ui.adsabs.harvard.edu/abs/2021A&A...649A...6G} {649, A6}

\bibitem[\protect\citeauthoryear{{Gaia Collaboration} et~al.,}{{Gaia
  Collaboration} et~al.}{2022}]{2022arXiv220800211G}
{Gaia Collaboration} et~al., 2022, arXiv e-prints, \href
  {https://ui.adsabs.harvard.edu/abs/2022arXiv220800211G} {p. arXiv:2208.00211}

\bibitem[\protect\citeauthoryear{{Ghodla}, {Eldridge}, {Stanway}  \&
  {Stevance}}{{Ghodla} et~al.}{2022}]{2022arXiv220803999G}
{Ghodla} S.,  {Eldridge} J.~J.,  {Stanway} E.~R.,   {Stevance} H.~F.,  2022,
  arXiv e-prints, \href {https://ui.adsabs.harvard.edu/abs/2022arXiv220803999G}
  {p. arXiv:2208.03999}

\bibitem[\protect\citeauthoryear{{Ghodla}, {Eldridge}, {Stanway}  \&
  {Stevance}}{{Ghodla} et~al.}{2023}]{2023MNRAS.518..860G}
{Ghodla} S.,  {Eldridge} J.~J.,  {Stanway} E.~R.,   {Stevance} H.~F.,  2023,
  \mn@doi [\mnras] {10.1093/mnras/stac3177}, \href
  {https://ui.adsabs.harvard.edu/abs/2023MNRAS.518..860G} {518, 860}

\bibitem[\protect\citeauthoryear{{Giacobbo} \& {Mapelli}}{{Giacobbo} \&
  {Mapelli}}{2018}]{2018MNRAS.480.2011G}
{Giacobbo} N.,  {Mapelli} M.,  2018, \mn@doi [\mnras] {10.1093/mnras/sty1999},
  \href {https://ui.adsabs.harvard.edu/abs/2018MNRAS.480.2011G} {480, 2011}

\bibitem[\protect\citeauthoryear{{Giacobbo} \& {Mapelli}}{{Giacobbo} \&
  {Mapelli}}{2019}]{2019MNRAS.482.2234G}
{Giacobbo} N.,  {Mapelli} M.,  2019, \mn@doi [\mnras] {10.1093/mnras/sty2848},
  \href {https://ui.adsabs.harvard.edu/abs/2019MNRAS.482.2234G} {482, 2234}

\bibitem[\protect\citeauthoryear{{Giacobbo} \& {Mapelli}}{{Giacobbo} \&
  {Mapelli}}{2020}]{2020ApJ...891..141G}
{Giacobbo} N.,  {Mapelli} M.,  2020, \mn@doi [\apj] {10.3847/1538-4357/ab7335},
  \href {https://ui.adsabs.harvard.edu/abs/2020ApJ...891..141G} {891, 141}

\bibitem[\protect\citeauthoryear{{Giardino} et~al.,}{{Giardino}
  et~al.}{2022}]{2022SPIE12180E..0XG}
{Giardino} G.,  et~al., 2022, in {Coyle} L.~E.,  {Matsuura} S.,   {Perrin}
  M.~D.,  eds,  Society of Photo-Optical Instrumentation Engineers (SPIE)
  Conference Series Vol. 12180, Space Telescopes and Instrumentation 2022:
  Optical, Infrared, and Millimeter Wave. p. 121800X (\mn@eprint {arXiv}
  {2208.04876}), \mn@doi{10.1117/12.2628980}

\bibitem[\protect\citeauthoryear{{Gies}}{{Gies}}{1987}]{1987ApJS...64..545G}
{Gies} D.~R.,  1987, \mn@doi [\apjs] {10.1086/191208}, \href
  {https://ui.adsabs.harvard.edu/abs/1987ApJS...64..545G} {64, 545}

\bibitem[\protect\citeauthoryear{{Gilkis} \& {Arcavi}}{{Gilkis} \&
  {Arcavi}}{2022}]{2022MNRAS.511..691G}
{Gilkis} A.,  {Arcavi} I.,  2022, \mn@doi [\mnras] {10.1093/mnras/stac088},
  \href {https://ui.adsabs.harvard.edu/abs/2022MNRAS.511..691G} {511, 691}

\bibitem[\protect\citeauthoryear{{Green}}{{Green}}{2014}]{2014BASI...42...47G}
{Green} D.~A.,  2014, Bulletin of the Astronomical Society of India, \href
  {https://ui.adsabs.harvard.edu/abs/2014BASI...42...47G} {42, 47}

\bibitem[\protect\citeauthoryear{{Green}}{{Green}}{2018}]{2018JOSS....3..695M}
{Green} G.,  2018, \mn@doi [The Journal of Open Source Software]
  {10.21105/joss.00695}, \href
  {https://ui.adsabs.harvard.edu/abs/2018JOSS....3..695M} {3, 695}

\bibitem[\protect\citeauthoryear{{Green}, {Schlafly}, {Zucker}, {Speagle}  \&
  {Finkbeiner}}{{Green} et~al.}{2019}]{2019ApJ...887...93G}
{Green} G.~M.,  {Schlafly} E.,  {Zucker} C.,  {Speagle} J.~S.,   {Finkbeiner}
  D.,  2019, \mn@doi [\apj] {10.3847/1538-4357/ab5362}, \href
  {https://ui.adsabs.harvard.edu/abs/2019ApJ...887...93G} {887, 93}

\bibitem[\protect\citeauthoryear{{Griggio}, {Nardiello}  \& {Bedin}}{{Griggio}
  et~al.}{2022}]{2022arXiv221203256G}
{Griggio} M.,  {Nardiello} D.,   {Bedin} L.~R.,  2022, arXiv e-prints, \href
  {https://ui.adsabs.harvard.edu/abs/2022arXiv221203256G} {p. arXiv:2212.03256}

\bibitem[\protect\citeauthoryear{{Guiglion} et~al.,}{{Guiglion}
  et~al.}{2015}]{2015A&A...583A..91G}
{Guiglion} G.,  et~al., 2015, \mn@doi [\aap] {10.1051/0004-6361/201525883},
  \href {https://ui.adsabs.harvard.edu/abs/2015A&A...583A..91G} {583, A91}

\bibitem[\protect\citeauthoryear{{Gvaramadze}, {R{\"o}ser}, {Scholz}  \&
  {Schilbach}}{{Gvaramadze} et~al.}{2011}]{2011A&A...529A..14G}
{Gvaramadze} V.~V.,  {R{\"o}ser} S.,  {Scholz} R.~D.,   {Schilbach} E.,  2011,
  \mn@doi [\aap] {10.1051/0004-6361/201016256}, \href
  {https://ui.adsabs.harvard.edu/abs/2011A&A...529A..14G} {529, A14}

\bibitem[\protect\citeauthoryear{{Gvaramadze}, {Weidner}, {Kroupa}  \&
  {Pflamm-Altenburg}}{{Gvaramadze} et~al.}{2012}]{2012MNRAS.424.3037G}
{Gvaramadze} V.~V.,  {Weidner} C.,  {Kroupa} P.,   {Pflamm-Altenburg} J.,
  2012, \mn@doi [\mnras] {10.1111/j.1365-2966.2012.21452.x}, \href
  {https://ui.adsabs.harvard.edu/abs/2012MNRAS.424.3037G} {424, 3037}

\bibitem[\protect\citeauthoryear{{Harris} et~al.,}{{Harris}
  et~al.}{2020}]{2020arXiv200610256H}
{Harris} C.~R.,  et~al., 2020, arXiv e-prints, \href
  {https://ui.adsabs.harvard.edu/abs/2020arXiv200610256H} {p. arXiv:2006.10256}

\bibitem[\protect\citeauthoryear{{Helfand}, {Chatterjee}, {Brisken}, {Camilo},
  {Reynolds}, {van Kerkwijk}, {Halpern}  \& {Ransom}}{{Helfand}
  et~al.}{2007}]{2007ApJ...662.1198H}
{Helfand} D.~J.,  {Chatterjee} S.,  {Brisken} W.~F.,  {Camilo} F.,  {Reynolds}
  J.,  {van Kerkwijk} M.~H.,  {Halpern} J.~P.,   {Ransom} S.~M.,  2007, \mn@doi
  [\apj] {10.1086/518028}, \href
  {https://ui.adsabs.harvard.edu/abs/2007ApJ...662.1198H} {662, 1198}

\bibitem[\protect\citeauthoryear{{Hirai}, {Podsiadlowski}  \& {Yamada}}{{Hirai}
  et~al.}{2018}]{2018ApJ...864..119H}
{Hirai} R.,  {Podsiadlowski} P.,   {Yamada} S.,  2018, \mn@doi [\apj]
  {10.3847/1538-4357/aad6a0}, \href
  {https://ui.adsabs.harvard.edu/abs/2018ApJ...864..119H} {864, 119}

\bibitem[\protect\citeauthoryear{{Hobbs}, {Lorimer}, {Lyne}  \&
  {Kramer}}{{Hobbs} et~al.}{2005}]{2005MNRAS.360..974H}
{Hobbs} G.,  {Lorimer} D.~R.,  {Lyne} A.~G.,   {Kramer} M.,  2005, \mn@doi
  [\mnras] {10.1111/j.1365-2966.2005.09087.x}, \href
  {https://ui.adsabs.harvard.edu/abs/2005MNRAS.360..974H} {360, 974}

\bibitem[\protect\citeauthoryear{{Hobbs} et~al.,}{{Hobbs}
  et~al.}{2016}]{2016arXiv160907325H}
{Hobbs} D.,  et~al., 2016, arXiv e-prints, \href
  {https://ui.adsabs.harvard.edu/abs/2016arXiv160907325H} {p. arXiv:1609.07325}

\bibitem[\protect\citeauthoryear{{Hunter}}{{Hunter}}{2007}]{2007CSE.....9...90H}
{Hunter} J.~D.,  2007, \mn@doi [Computing in Science and Engineering]
  {10.1109/MCSE.2007.55}, \href
  {https://ui.adsabs.harvard.edu/abs/2007CSE.....9...90H} {9, 90}

\bibitem[\protect\citeauthoryear{{Igoshev}}{{Igoshev}}{2020}]{2020MNRAS.494.3663I}
{Igoshev} A.~P.,  2020, \mn@doi [\mnras] {10.1093/mnras/staa958}, \href
  {https://ui.adsabs.harvard.edu/abs/2020MNRAS.494.3663I} {494, 3663}

\bibitem[\protect\citeauthoryear{{Igoshev} \& {Perets}}{{Igoshev} \&
  {Perets}}{2019}]{2019MNRAS.486.4098I}
{Igoshev} A.~P.,  {Perets} H.~B.,  2019, \mn@doi [\mnras]
  {10.1093/mnras/stz1024}, \href
  {https://ui.adsabs.harvard.edu/abs/2019MNRAS.486.4098I} {486, 4098}

\bibitem[\protect\citeauthoryear{{Igoshev}, {Chruslinska}, {Dorozsmai}  \&
  {Toonen}}{{Igoshev} et~al.}{2021}]{2021MNRAS.508.3345I}
{Igoshev} A.~P.,  {Chruslinska} M.,  {Dorozsmai} A.,   {Toonen} S.,  2021,
  \mn@doi [\mnras] {10.1093/mnras/stab2734}, \href
  {https://ui.adsabs.harvard.edu/abs/2021MNRAS.508.3345I} {508, 3345}

\bibitem[\protect\citeauthoryear{{Iorio} et~al.,}{{Iorio}
  et~al.}{2022}]{2022arXiv221111774I}
{Iorio} G.,  et~al., 2022, arXiv e-prints, \href
  {https://ui.adsabs.harvard.edu/abs/2022arXiv221111774I} {p. arXiv:2211.11774}

\bibitem[\protect\citeauthoryear{{Janka}}{{Janka}}{2012}]{2012ARNPS..62..407J}
{Janka} H.-T.,  2012, \mn@doi [Annual Review of Nuclear and Particle Science]
  {10.1146/annurev-nucl-102711-094901}, \href
  {https://ui.adsabs.harvard.edu/abs/2012ARNPS..62..407J} {62, 407}

\bibitem[\protect\citeauthoryear{{Janka}}{{Janka}}{2017}]{2017ApJ...837...84J}
{Janka} H.-T.,  2017, \mn@doi [\apj] {10.3847/1538-4357/aa618e}, \href
  {https://ui.adsabs.harvard.edu/abs/2017ApJ...837...84J} {837, 84}

\bibitem[\protect\citeauthoryear{{Kalari}, {Vink}, {de Wit}, {Bastian}  \&
  {M{\'e}ndez}}{{Kalari} et~al.}{2019}]{2019A&A...625L...2K}
{Kalari} V.~M.,  {Vink} J.~S.,  {de Wit} W.~J.,  {Bastian} N.~J.,
  {M{\'e}ndez} R.~A.,  2019, \mn@doi [\aap] {10.1051/0004-6361/201935107},
  \href {https://ui.adsabs.harvard.edu/abs/2019A&A...625L...2K} {625, L2}

\bibitem[\protect\citeauthoryear{{Kapil}, {Mandel}, {Berti}  \&
  {M{\"u}ller}}{{Kapil} et~al.}{2022}]{2022arXiv220909252K}
{Kapil} V.,  {Mandel} I.,  {Berti} E.,   {M{\"u}ller} B.,  2022, arXiv
  e-prints, \href {https://ui.adsabs.harvard.edu/abs/2022arXiv220909252K} {p.
  arXiv:2209.09252}

\bibitem[\protect\citeauthoryear{{Kerzendorf} et~al.,}{{Kerzendorf}
  et~al.}{2019}]{2019AA...623A..34K}
{Kerzendorf} W.~E.,  et~al., 2019, \mn@doi [\aap]
  {10.1051/0004-6361/201732206}, \href
  {https://ui.adsabs.harvard.edu/abs/2019A&A...623A..34K} {623, A34}

\bibitem[\protect\citeauthoryear{{Kochanek}}{{Kochanek}}{2015}]{2015MNRAS.446.1213K}
{Kochanek} C.~S.,  2015, \mn@doi [\mnras] {10.1093/mnras/stu2056}, \href
  {https://ui.adsabs.harvard.edu/abs/2015MNRAS.446.1213K} {446, 1213}

\bibitem[\protect\citeauthoryear{{Kochanek}}{{Kochanek}}{2018}]{2018MNRAS.473.1633K}
{Kochanek} C.~S.,  2018, \mn@doi [\mnras] {10.1093/mnras/stx2423}, \href
  {https://ui.adsabs.harvard.edu/abs/2018MNRAS.473.1633K} {473, 1633}

\bibitem[\protect\citeauthoryear{{Laplace}, {G{\"o}tberg}, {de Mink}, {Justham}
   \& {Farmer}}{{Laplace} et~al.}{2020}]{2020A&A...637A...6L}
{Laplace} E.,  {G{\"o}tberg} Y.,  {de Mink} S.~E.,  {Justham} S.,   {Farmer}
  R.,  2020, \mn@doi [\aap] {10.1051/0004-6361/201937300}, \href
  {https://ui.adsabs.harvard.edu/abs/2020A&A...637A...6L} {637, A6}

\bibitem[\protect\citeauthoryear{{Laplace}, {Justham}, {Renzo}, {G{\"o}tberg},
  {Farmer}, {Vartanyan}  \& {de Mink}}{{Laplace}
  et~al.}{2021}]{2021A&A...656A..58L}
{Laplace} E.,  {Justham} S.,  {Renzo} M.,  {G{\"o}tberg} Y.,  {Farmer} R.,
  {Vartanyan} D.,   {de Mink} S.~E.,  2021, \mn@doi [\aap]
  {10.1051/0004-6361/202140506}, \href
  {https://ui.adsabs.harvard.edu/abs/2021A&A...656A..58L} {656, A58}

\bibitem[\protect\citeauthoryear{{Lattimer}}{{Lattimer}}{2012}]{2012ARNPS..62..485L}
{Lattimer} J.~M.,  2012, \mn@doi [Annual Review of Nuclear and Particle
  Science] {10.1146/annurev-nucl-102711-095018}, \href
  {https://ui.adsabs.harvard.edu/abs/2012ARNPS..62..485L} {62, 485}

\bibitem[\protect\citeauthoryear{{Laureijs} et~al.,}{{Laureijs}
  et~al.}{2011}]{2011arXiv1110.3193L}
{Laureijs} R.,  et~al., 2011, arXiv e-prints, \href
  {https://ui.adsabs.harvard.edu/abs/2011arXiv1110.3193L} {p. arXiv:1110.3193}

\bibitem[\protect\citeauthoryear{{Lennon} et~al.,}{{Lennon}
  et~al.}{2018}]{2018A&A...619A..78L}
{Lennon} D.~J.,  et~al., 2018, \mn@doi [\aap] {10.1051/0004-6361/201833465},
  \href {https://ui.adsabs.harvard.edu/abs/2018A&A...619A..78L} {619, A78}

\bibitem[\protect\citeauthoryear{{Lindegren} et~al.,}{{Lindegren}
  et~al.}{2021}]{2021A&A...649A...2L}
{Lindegren} L.,  et~al., 2021, \mn@doi [\aap] {10.1051/0004-6361/202039709},
  \href {https://ui.adsabs.harvard.edu/abs/2021A&A...649A...2L} {649, A2}

\bibitem[\protect\citeauthoryear{{Liu}, {Tauris}, {R{\"o}pke}, {Moriya},
  {Kruckow}, {Stancliffe}  \& {Izzard}}{{Liu}
  et~al.}{2015}]{2015A&A...584A..11L}
{Liu} Z.-W.,  {Tauris} T.~M.,  {R{\"o}pke} F.~K.,  {Moriya} T.~J.,  {Kruckow}
  M.,  {Stancliffe} R.~J.,   {Izzard} R.~G.,  2015, \mn@doi [\aap]
  {10.1051/0004-6361/201526757}, \href
  {https://ui.adsabs.harvard.edu/abs/2015A&A...584A..11L} {584, A11}

\bibitem[\protect\citeauthoryear{{Lux}, {Neuh{\"a}user}, {Mugrauer}  \&
  {Bischoff}}{{Lux} et~al.}{2021}]{2021AN....342..553L}
{Lux} O.,  {Neuh{\"a}user} R.,  {Mugrauer} M.,   {Bischoff} R.,  2021, \mn@doi
  [Astronomische Nachrichten] {10.1002/asna.202113860}, \href
  {https://ui.adsabs.harvard.edu/abs/2021AN....342..553L} {342, 553}

\bibitem[\protect\citeauthoryear{{Lyman}, {Levan}, {Wiersema}, {Kouveliotou},
  {Chrimes}  \& {Fruchter}}{{Lyman} et~al.}{2022}]{2022ApJ...926..121L}
{Lyman} J.~D.,  {Levan} A.~J.,  {Wiersema} K.,  {Kouveliotou} C.,  {Chrimes}
  A.~A.,   {Fruchter} A.~S.,  2022, \mn@doi [\apj] {10.3847/1538-4357/ac432f},
  \href {https://ui.adsabs.harvard.edu/abs/2022ApJ...926..121L} {926, 121}

\bibitem[\protect\citeauthoryear{{Ma{\'\i}z Apell{\'a}niz}, {Pantaleoni
  Gonz{\'a}lez}, {Barb{\'a}}, {Sim{\'o}n-D{\'\i}az}, {Negueruela}, {Lennon},
  {Sota}  \& {Trigueros P{\'a}ez}}{{Ma{\'\i}z Apell{\'a}niz}
  et~al.}{2018}]{2018A&A...616A.149M}
{Ma{\'\i}z Apell{\'a}niz} J.,  {Pantaleoni Gonz{\'a}lez} M.,  {Barb{\'a}}
  R.~H.,  {Sim{\'o}n-D{\'\i}az} S.,  {Negueruela} I.,  {Lennon} D.~J.,  {Sota}
  A.,   {Trigueros P{\'a}ez} E.,  2018, \mn@doi [\aap]
  {10.1051/0004-6361/201832787}, \href
  {https://ui.adsabs.harvard.edu/abs/2018A&A...616A.149M} {616, A149}

\bibitem[\protect\citeauthoryear{{Manchester}, {Hobbs}, {Teoh}  \&
  {Hobbs}}{{Manchester} et~al.}{2005}]{2005AJ....129.1993M}
{Manchester} R.~N.,  {Hobbs} G.~B.,  {Teoh} A.,   {Hobbs} M.,  2005, \mn@doi
  [\aj] {10.1086/428488}, \href
  {https://ui.adsabs.harvard.edu/abs/2005AJ....129.1993M} {129, 1993}

\bibitem[\protect\citeauthoryear{{Mandel} \& {de Mink}}{{Mandel} \& {de
  Mink}}{2016}]{2016MNRAS.458.2634M}
{Mandel} I.,  {de Mink} S.~E.,  2016, \mn@doi [\mnras] {10.1093/mnras/stw379},
  \href {https://ui.adsabs.harvard.edu/abs/2016MNRAS.458.2634M} {458, 2634}

\bibitem[\protect\citeauthoryear{{Marchant}, {Pappas}, {Gallegos-Garcia},
  {Berry}, {Taam}, {Kalogera}  \& {Podsiadlowski}}{{Marchant}
  et~al.}{2021}]{2021A&A...650A.107M}
{Marchant} P.,  {Pappas} K. M.~W.,  {Gallegos-Garcia} M.,  {Berry} C. P.~L.,
  {Taam} R.~E.,  {Kalogera} V.,   {Podsiadlowski} P.,  2021, \mn@doi [\aap]
  {10.1051/0004-6361/202039992}, \href
  {https://ui.adsabs.harvard.edu/abs/2021A&A...650A.107M} {650, A107}

\bibitem[\protect\citeauthoryear{{Marchetti}, {Evans}  \& {Rossi}}{{Marchetti}
  et~al.}{2022}]{2022MNRAS.515..767M}
{Marchetti} T.,  {Evans} F.~A.,   {Rossi} E.~M.,  2022, \mn@doi [\mnras]
  {10.1093/mnras/stac1777}, \href
  {https://ui.adsabs.harvard.edu/abs/2022MNRAS.515..767M} {515, 767}

\bibitem[\protect\citeauthoryear{{Moe} \& {Di Stefano}}{{Moe} \& {Di
  Stefano}}{2017}]{2017ApJS..230...15M}
{Moe} M.,  {Di Stefano} R.,  2017, \mn@doi [\apjs] {10.3847/1538-4365/aa6fb6},
  \href {https://ui.adsabs.harvard.edu/abs/2017ApJS..230...15M} {230, 15}

\bibitem[\protect\citeauthoryear{{Nelemans} \& {Tout}}{{Nelemans} \&
  {Tout}}{2005}]{2005MNRAS.356..753N}
{Nelemans} G.,  {Tout} C.~A.,  2005, \mn@doi [\mnras]
  {10.1111/j.1365-2966.2004.08496.x}, \href
  {https://ui.adsabs.harvard.edu/abs/2005MNRAS.356..753N} {356, 753}

\bibitem[\protect\citeauthoryear{{Nelemans}, {Verbunt}, {Yungelson}  \&
  {Portegies Zwart}}{{Nelemans} et~al.}{2000}]{2000A&A...360.1011N}
{Nelemans} G.,  {Verbunt} F.,  {Yungelson} L.~R.,   {Portegies Zwart} S.~F.,
  2000, \mn@doi [\aap] {10.48550/arXiv.astro-ph/0006216}, \href
  {https://ui.adsabs.harvard.edu/abs/2000A&A...360.1011N} {360, 1011}

\bibitem[\protect\citeauthoryear{{Neuh{\"a}user}, {Gie{\ss}ler}  \&
  {Hambaryan}}{{Neuh{\"a}user} et~al.}{2020}]{2020MNRAS.498..899N}
{Neuh{\"a}user} R.,  {Gie{\ss}ler} F.,   {Hambaryan} V.~V.,  2020, \mn@doi
  [\mnras] {10.1093/mnras/stz2629}, \href
  {https://ui.adsabs.harvard.edu/abs/2020MNRAS.498..899N} {498, 899}

\bibitem[\protect\citeauthoryear{{Nomoto}, {Iwamoto}  \& {Suzuki}}{{Nomoto}
  et~al.}{1995}]{1995PhR...256..173N}
{Nomoto} K.~I.,  {Iwamoto} K.,   {Suzuki} T.,  1995, \mn@doi [\physrep]
  {10.1016/0370-1573(94)00107-E}, \href
  {https://ui.adsabs.harvard.edu/abs/1995PhR...256..173N} {256, 173}

\bibitem[\protect\citeauthoryear{{Nugis} \& {Lamers}}{{Nugis} \&
  {Lamers}}{2000}]{2000A&A...360..227N}
{Nugis} T.,  {Lamers} H.~J.~G.~L.~M.,  2000, \aap, \href
  {https://ui.adsabs.harvard.edu/abs/2000A&A...360..227N} {360, 227}

\bibitem[\protect\citeauthoryear{{Oey} et~al.,}{{Oey}
  et~al.}{2018}]{2018ApJ...867L...8O}
{Oey} M.~S.,  et~al., 2018, \mn@doi [\apjl] {10.3847/2041-8213/aae892}, \href
  {https://ui.adsabs.harvard.edu/abs/2018ApJ...867L...8O} {867, L8}

\bibitem[\protect\citeauthoryear{{Offner}, {Moe}, {Kratter}, {Sadavoy},
  {Jensen}  \& {Tobin}}{{Offner} et~al.}{2022}]{2022arXiv220310066O}
{Offner} S. S.~R.,  {Moe} M.,  {Kratter} K.~M.,  {Sadavoy} S.~I.,  {Jensen} E.
  L.~N.,   {Tobin} J.~J.,  2022, \mn@doi [arXiv e-prints]
  {10.48550/arXiv.2203.10066}, \href
  {https://ui.adsabs.harvard.edu/abs/2022arXiv220310066O} {p. arXiv:2203.10066}

\bibitem[\protect\citeauthoryear{{Ogata}, {Hirai}  \& {Hijikawa}}{{Ogata}
  et~al.}{2021}]{2021MNRAS.505.2485O}
{Ogata} M.,  {Hirai} R.,   {Hijikawa} K.,  2021, \mn@doi [\mnras]
  {10.1093/mnras/stab1439}, \href
  {https://ui.adsabs.harvard.edu/abs/2021MNRAS.505.2485O} {505, 2485}

\bibitem[\protect\citeauthoryear{{Oke} \& {Gunn}}{{Oke} \&
  {Gunn}}{1983}]{1983ApJ...266..713O}
{Oke} J.~B.,  {Gunn} J.~E.,  1983, \mn@doi [\apj] {10.1086/160817}, \href
  {http://adsabs.harvard.edu/abs/1983ApJ...266..713O} {266, 713}

\bibitem[\protect\citeauthoryear{{Olausen} \& {Kaspi}}{{Olausen} \&
  {Kaspi}}{2014}]{2014ApJS..212....6O}
{Olausen} S.~A.,  {Kaspi} V.~M.,  2014, \mn@doi [\apjs]
  {10.1088/0067-0049/212/1/6}, \href
  {https://ui.adsabs.harvard.edu/abs/2014ApJS..212....6O} {212, 6}

\bibitem[\protect\citeauthoryear{{Patton} \& {Sukhbold}}{{Patton} \&
  {Sukhbold}}{2020}]{2020MNRAS.499.2803P}
{Patton} R.~A.,  {Sukhbold} T.,  2020, \mn@doi [\mnras]
  {10.1093/mnras/staa3029}, \href
  {https://ui.adsabs.harvard.edu/abs/2020MNRAS.499.2803P} {499, 2803}

\bibitem[\protect\citeauthoryear{{Pavlovskii}, {Ivanova}, {Belczynski}  \&
  {Van}}{{Pavlovskii} et~al.}{2017}]{2017MNRAS.465.2092P}
{Pavlovskii} K.,  {Ivanova} N.,  {Belczynski} K.,   {Van} K.~X.,  2017, \mn@doi
  [\mnras] {10.1093/mnras/stw2786}, \href
  {https://ui.adsabs.harvard.edu/abs/2017MNRAS.465.2092P} {465, 2092}

\bibitem[\protect\citeauthoryear{{Perets} \& {Beniamini}}{{Perets} \&
  {Beniamini}}{2021}]{2021MNRAS.503.5997P}
{Perets} H.~B.,  {Beniamini} P.,  2021, \mn@doi [\mnras]
  {10.1093/mnras/stab794}, \href
  {https://ui.adsabs.harvard.edu/abs/2021MNRAS.503.5997P} {503, 5997}

\bibitem[\protect\citeauthoryear{{Perez} \& {Granger}}{{Perez} \&
  {Granger}}{2007}]{2007CSE.....9c..21P}
{Perez} F.,  {Granger} B.~E.,  2007, \mn@doi [Computing in Science and
  Engineering] {10.1109/MCSE.2007.53}, \href
  {https://ui.adsabs.harvard.edu/abs/2007CSE.....9c..21P} {9, 21}

\bibitem[\protect\citeauthoryear{{Pflamm-Altenburg} \&
  {Kroupa}}{{Pflamm-Altenburg} \& {Kroupa}}{2010}]{2010MNRAS.404.1564P}
{Pflamm-Altenburg} J.,  {Kroupa} P.,  2010, \mn@doi [\mnras]
  {10.1111/j.1365-2966.2010.16376.x}, \href
  {https://ui.adsabs.harvard.edu/abs/2010MNRAS.404.1564P} {404, 1564}

\bibitem[\protect\citeauthoryear{{Podsiadlowski}, {Joss}  \&
  {Hsu}}{{Podsiadlowski} et~al.}{1992}]{1992ApJ...391..246P}
{Podsiadlowski} P.,  {Joss} P.~C.,   {Hsu} J.~J.~L.,  1992, \mn@doi [\apj]
  {10.1086/171341}, \href
  {https://ui.adsabs.harvard.edu/abs/1992ApJ...391..246P} {391, 246}

\bibitem[\protect\citeauthoryear{{Pols}, {Cote}, {Waters}  \& {Heise}}{{Pols}
  et~al.}{1991}]{1991A&A...241..419P}
{Pols} O.~R.,  {Cote} J.,  {Waters} L.~B.~F.~M.,   {Heise} J.,  1991, \aap,
  \href {https://ui.adsabs.harvard.edu/abs/1991A&A...241..419P} {241, 419}

\bibitem[\protect\citeauthoryear{{Poveda}, {Ruiz}  \& {Allen}}{{Poveda}
  et~al.}{1967}]{1967BOTT....4...86P}
{Poveda} A.,  {Ruiz} J.,   {Allen} C.,  1967, Boletin de los Observatorios
  Tonantzintla y Tacubaya, \href
  {https://ui.adsabs.harvard.edu/abs/1967BOTT....4...86P} {4, 86}

\bibitem[\protect\citeauthoryear{{Price-Whelan} et~al.,}{{Price-Whelan}
  et~al.}{2018}]{astropy:2018}
{Price-Whelan} A.~M.,  et~al., 2018, \mn@doi [\aj] {10.3847/1538-3881/aabc4f},
  \href {https://ui.adsabs.harvard.edu/#abs/2018AJ....156..123T} {156, 123}

\bibitem[\protect\citeauthoryear{{Renzo} \& {G{\"o}tberg}}{{Renzo} \&
  {G{\"o}tberg}}{2021}]{2021ApJ...923..277R}
{Renzo} M.,  {G{\"o}tberg} Y.,  2021, \mn@doi [\apj]
  {10.3847/1538-4357/ac29c5}, \href
  {https://ui.adsabs.harvard.edu/abs/2021ApJ...923..277R} {923, 277}

\bibitem[\protect\citeauthoryear{{Renzo} et~al.,}{{Renzo}
  et~al.}{2019}]{2019A&A...624A..66R}
{Renzo} M.,  et~al., 2019, \mn@doi [\aap] {10.1051/0004-6361/201833297}, \href
  {https://ui.adsabs.harvard.edu/abs/2019A&A...624A..66R} {624, A66}

\bibitem[\protect\citeauthoryear{{Renzo}, {Zapartas}, {Justham}, {Breivik},
  {Lau}, {Farmer}, {Cantiello}  \& {Metzger}}{{Renzo}
  et~al.}{2023}]{2023ApJ...942L..32R}
{Renzo} M.,  {Zapartas} E.,  {Justham} S.,  {Breivik} K.,  {Lau} M.,  {Farmer}
  R.,  {Cantiello} M.,   {Metzger} B.~D.,  2023, \mn@doi [\apjl]
  {10.3847/2041-8213/aca4d3}, \href
  {https://ui.adsabs.harvard.edu/abs/2023ApJ...942L..32R} {942, L32}

\bibitem[\protect\citeauthoryear{{Richards}, {Eldridge}, {Briel}, {Stevance}
  \& {Willcox}}{{Richards} et~al.}{2022}]{2022arXiv220802407R}
{Richards} S.~M.,  {Eldridge} J.~J.,  {Briel} M.~M.,  {Stevance} H.~F.,
  {Willcox} R.,  2022, arXiv e-prints, \href
  {https://ui.adsabs.harvard.edu/abs/2022arXiv220802407R} {p. arXiv:2208.02407}

\bibitem[\protect\citeauthoryear{{Riello} et~al.,}{{Riello}
  et~al.}{2021}]{2021A&A...649A...3R}
{Riello} M.,  et~al., 2021, \mn@doi [\aap] {10.1051/0004-6361/202039587}, \href
  {https://ui.adsabs.harvard.edu/abs/2021A&A...649A...3R} {649, A3}

\bibitem[\protect\citeauthoryear{{Rodrigo} \& {Solano}}{{Rodrigo} \&
  {Solano}}{2020}]{2020sea..confE.182R}
{Rodrigo} C.,  {Solano} E.,  2020, in XIV.0 Scientific Meeting (virtual) of the
  Spanish Astronomical Society. p.~182

\bibitem[\protect\citeauthoryear{{Rodrigo}, {Solano}  \& {Bayo}}{{Rodrigo}
  et~al.}{2012}]{2012ivoa.rept.1015R}
{Rodrigo} C.,  {Solano} E.,   {Bayo} A.,  2012, {SVO Filter Profile Service
  Version 1.0}, IVOA Working Draft 15 October 2012,
  \mn@doi{10.5479/ADS/bib/2012ivoa.rept.1015R}

\bibitem[\protect\citeauthoryear{{Sana} et~al.,}{{Sana}
  et~al.}{2012}]{2012Sci...337..444S}
{Sana} H.,  et~al., 2012, \mn@doi [Science] {10.1126/science.1223344}, \href
  {https://ui.adsabs.harvard.edu/abs/2012Sci...337..444S} {337, 444}

\bibitem[\protect\citeauthoryear{{Sana} et~al.,}{{Sana}
  et~al.}{2014}]{2014ApJS..215...15S}
{Sana} H.,  et~al., 2014, \mn@doi [\apjs] {10.1088/0067-0049/215/1/15}, \href
  {https://ui.adsabs.harvard.edu/abs/2014ApJS..215...15S} {215, 15}

\bibitem[\protect\citeauthoryear{{Sana} et~al.,}{{Sana}
  et~al.}{2022}]{2022arXiv221113476S}
{Sana} H.,  et~al., 2022, arXiv e-prints, \href
  {https://ui.adsabs.harvard.edu/abs/2022arXiv221113476S} {p. arXiv:2211.13476}

\bibitem[\protect\citeauthoryear{{Stanway} \& {Eldridge}}{{Stanway} \&
  {Eldridge}}{2018}]{2018MNRAS.479...75S}
{Stanway} E.~R.,  {Eldridge} J.~J.,  2018, \mn@doi [\mnras]
  {10.1093/mnras/sty1353}, \href
  {https://ui.adsabs.harvard.edu/abs/2018MNRAS.479...75S} {479, 75}

\bibitem[\protect\citeauthoryear{{Stanway}, {Chrimes}, {Eldridge}  \&
  {Stevance}}{{Stanway} et~al.}{2020}]{2020MNRAS.495.4605S}
{Stanway} E.~R.,  {Chrimes} A.~A.,  {Eldridge} J.~J.,   {Stevance} H.~F.,
  2020, \mn@doi [\mnras] {10.1093/mnras/staa1166}, \href
  {https://ui.adsabs.harvard.edu/abs/2020MNRAS.495.4605S} {495, 4605}

\bibitem[\protect\citeauthoryear{{Stevance}, {Eldridge}  \&
  {Stanway}}{{Stevance} et~al.}{2020}]{2020JOSS....5.1987S}
{Stevance} H.,  {Eldridge} J.,   {Stanway} E.,  2020, \mn@doi [The Journal of
  Open Source Software] {10.21105/joss.01987}, \href
  {https://ui.adsabs.harvard.edu/abs/2020JOSS....5.1987S} {5, 1987}

\bibitem[\protect\citeauthoryear{{Stone}}{{Stone}}{1991}]{1991AJ....102..333S}
{Stone} R.~C.,  1991, \mn@doi [\aj] {10.1086/115880}, \href
  {https://ui.adsabs.harvard.edu/abs/1991AJ....102..333S} {102, 333}

\bibitem[\protect\citeauthoryear{{Sun}, {Maund}  \& {Crowther}}{{Sun}
  et~al.}{2022}]{2022arXiv220905283S}
{Sun} N.-C.,  {Maund} J.~R.,   {Crowther} P.~A.,  2022, arXiv e-prints, \href
  {https://ui.adsabs.harvard.edu/abs/2022arXiv220905283S} {p. arXiv:2209.05283}

\bibitem[\protect\citeauthoryear{{Tauris} \& {Takens}}{{Tauris} \&
  {Takens}}{1998}]{1998A&A...330.1047T}
{Tauris} T.~M.,  {Takens} R.~J.,  1998, \aap, \href
  {https://ui.adsabs.harvard.edu/abs/1998A&A...330.1047T} {330, 1047}

\bibitem[\protect\citeauthoryear{{Tauris}, {Fender}, {van den Heuvel},
  {Johnston}  \& {Wu}}{{Tauris} et~al.}{1999}]{1999MNRAS.310.1165T}
{Tauris} T.~M.,  {Fender} R.~P.,  {van den Heuvel} E.~P.~J.,  {Johnston} H.~M.,
    {Wu} K.,  1999, \mn@doi [\mnras] {10.1046/j.1365-8711.1999.03068.x}, \href
  {https://ui.adsabs.harvard.edu/abs/1999MNRAS.310.1165T} {310, 1165}

\bibitem[\protect\citeauthoryear{{Tauris} et~al.,}{{Tauris}
  et~al.}{2017}]{2017ApJ...846..170T}
{Tauris} T.~M.,  et~al., 2017, \mn@doi [\apj] {10.3847/1538-4357/aa7e89}, \href
  {https://ui.adsabs.harvard.edu/abs/2017ApJ...846..170T} {846, 170}

\bibitem[\protect\citeauthoryear{{Taylor} \& {Cordes}}{{Taylor} \&
  {Cordes}}{1993}]{1993ApJ...411..674T}
{Taylor} J.~H.,  {Cordes} J.~M.,  1993, \mn@doi [\apj] {10.1086/172870}, \href
  {https://ui.adsabs.harvard.edu/abs/1993ApJ...411..674T} {411, 674}

\bibitem[\protect\citeauthoryear{{Temmink}, {Pols}, {Justham}, {Istrate}  \&
  {Toonen}}{{Temmink} et~al.}{2023}]{2023A&A...669A..45T}
{Temmink} K.~D.,  {Pols} O.~R.,  {Justham} S.,  {Istrate} A.~G.,   {Toonen} S.,
   2023, \mn@doi [\aap] {10.1051/0004-6361/202244137}, \href
  {https://ui.adsabs.harvard.edu/abs/2023A&A...669A..45T} {669, A45}

\bibitem[\protect\citeauthoryear{{Tendulkar}, {Cameron}  \&
  {Kulkarni}}{{Tendulkar} et~al.}{2012}]{2012ApJ...761...76T}
{Tendulkar} S.~P.,  {Cameron} P.~B.,   {Kulkarni} S.~R.,  2012, \mn@doi [\apj]
  {10.1088/0004-637X/761/1/76}, \href
  {https://ui.adsabs.harvard.edu/abs/2012ApJ...761...76T} {761, 76}

\bibitem[\protect\citeauthoryear{{Tendulkar}, {Cameron}  \&
  {Kulkarni}}{{Tendulkar} et~al.}{2013}]{2013ApJ...772...31T}
{Tendulkar} S.~P.,  {Cameron} P.~B.,   {Kulkarni} S.~R.,  2013, \mn@doi [\apj]
  {10.1088/0004-637X/772/1/31}, \href
  {https://ui.adsabs.harvard.edu/abs/2013ApJ...772...31T} {772, 31}

\bibitem[\protect\citeauthoryear{{Tetzlaff}, {Neuh{\"a}user}  \&
  {Hohle}}{{Tetzlaff} et~al.}{2011}]{2011MNRAS.410..190T}
{Tetzlaff} N.,  {Neuh{\"a}user} R.,   {Hohle} M.~M.,  2011, \mn@doi [\mnras]
  {10.1111/j.1365-2966.2010.17434.x}, \href
  {https://ui.adsabs.harvard.edu/abs/2011MNRAS.410..190T} {410, 190}

\bibitem[\protect\citeauthoryear{{Verbunt}, {Igoshev}  \& {Cator}}{{Verbunt}
  et~al.}{2017}]{2017A&A...608A..57V}
{Verbunt} F.,  {Igoshev} A.,   {Cator} E.,  2017, \mn@doi [\aap]
  {10.1051/0004-6361/201731518}, \href
  {https://ui.adsabs.harvard.edu/abs/2017A&A...608A..57V} {608, A57}

\bibitem[\protect\citeauthoryear{{Vink}}{{Vink}}{2012}]{2012A&ARv..20...49V}
{Vink} J.,  2012, \mn@doi [\aapr] {10.1007/s00159-011-0049-1}, \href
  {https://ui.adsabs.harvard.edu/abs/2012A&ARv..20...49V} {20, 49}

\bibitem[\protect\citeauthoryear{{Vink}, {de Koter}  \& {Lamers}}{{Vink}
  et~al.}{2001}]{2001A&A...369..574V}
{Vink} J.~S.,  {de Koter} A.,   {Lamers} H.~J.~G.~L.~M.,  2001, \mn@doi [\aap]
  {10.1051/0004-6361:20010127}, \href
  {https://ui.adsabs.harvard.edu/abs/2001A&A...369..574V} {369, 574}

\bibitem[\protect\citeauthoryear{{Virtanen} et~al.,}{{Virtanen}
  et~al.}{2020}]{2020NatMe..17..261V}
{Virtanen} P.,  et~al., 2020, \mn@doi [Nature Methods]
  {10.1038/s41592-019-0686-2}, \href
  {https://ui.adsabs.harvard.edu/abs/2020NatMe..17..261V} {17, 261}

\bibitem[\protect\citeauthoryear{{WFIRST Astrometry Working Group}
  et~al.,}{{WFIRST Astrometry Working Group}
  et~al.}{2019}]{2019JATIS...5d4005W}
{WFIRST Astrometry Working Group} et~al., 2019, \mn@doi [Journal of
  Astronomical Telescopes, Instruments, and Systems]
  {10.1117/1.JATIS.5.4.044005}, \href
  {https://ui.adsabs.harvard.edu/abs/2019JATIS...5d4005W} {5, 044005}

\bibitem[\protect\citeauthoryear{{Wang}, {Zhang}, {Jiang}, {Zhao}, {Chen},
  {Chen}, {Gao}  \& {Liu}}{{Wang} et~al.}{2020}]{2020AA...639A..72W}
{Wang} S.,  {Zhang} C.,  {Jiang} B.,  {Zhao} H.,  {Chen} B.,  {Chen} X.,  {Gao}
  J.,   {Liu} J.,  2020, \mn@doi [\aap] {10.1051/0004-6361/201936868}, \href
  {https://ui.adsabs.harvard.edu/abs/2020A&A...639A..72W} {639, A72}

\bibitem[\protect\citeauthoryear{{Wang}, {Gies}, {Peters}, {G{\"o}tberg},
  {Chojnowski}, {Lester}  \& {Howell}}{{Wang}
  et~al.}{2021}]{2021AJ....161..248W}
{Wang} L.,  {Gies} D.~R.,  {Peters} G.~J.,  {G{\"o}tberg} Y.,  {Chojnowski}
  S.~D.,  {Lester} K.~V.,   {Howell} S.~B.,  2021, \mn@doi [\aj]
  {10.3847/1538-3881/abf144}, \href
  {https://ui.adsabs.harvard.edu/abs/2021AJ....161..248W} {161, 248}

\bibitem[\protect\citeauthoryear{Waskom}{Waskom}{2021}]{Waskom2021}
Waskom M.~L.,  2021, \mn@doi [Journal of Open Source Software]
  {10.21105/joss.03021}, 6, 3021

\bibitem[\protect\citeauthoryear{{Willcox}, {Mandel}, {Thrane}, {Deller},
  {Stevenson}  \& {Vigna-G{\'o}mez}}{{Willcox}
  et~al.}{2021}]{2021ApJ...920L..37W}
{Willcox} R.,  {Mandel} I.,  {Thrane} E.,  {Deller} A.,  {Stevenson} S.,
  {Vigna-G{\'o}mez} A.,  2021, \mn@doi [\apjl] {10.3847/2041-8213/ac2cc8},
  \href {https://ui.adsabs.harvard.edu/abs/2021ApJ...920L..37W} {920, L37}

\bibitem[\protect\citeauthoryear{{Zapartas} et~al.,}{{Zapartas}
  et~al.}{2021}]{2021A&A...656L..19Z}
{Zapartas} E.,  et~al., 2021, \mn@doi [\aap] {10.1051/0004-6361/202141506},
  \href {https://ui.adsabs.harvard.edu/abs/2021A&A...656L..19Z} {656, L19}

\bibitem[\protect\citeauthoryear{{de Jager}, {Nieuwenhuijzen}  \& {van der
  Hucht}}{{de Jager} et~al.}{1988}]{1988A&AS...72..259D}
{de Jager} C.,  {Nieuwenhuijzen} H.,   {van der Hucht} K.~A.,  1988, \aaps,
  \href {https://ui.adsabs.harvard.edu/abs/1988A&AS...72..259D} {72, 259}

\bibitem[\protect\citeauthoryear{{de Jong} et~al.,}{{de Jong}
  et~al.}{2014}]{2014SPIE.9147E..0MD}
{de Jong} R.~S.,  et~al., 2014, in {Ramsay} S.~K.,  {McLean} I.~S.,   {Takami}
  H.,  eds,  Society of Photo-Optical Instrumentation Engineers (SPIE)
  Conference Series Vol. 9147, Ground-based and Airborne Instrumentation for
  Astronomy V. p. 91470M, \mn@doi{10.1117/12.2055826}

\bibitem[\protect\citeauthoryear{{de Jong} et~al.,}{{de Jong}
  et~al.}{2019}]{2019Msngr.175....3D}
{de Jong} R.~S.,  et~al., 2019, \mn@doi [The Messenger]
  {10.18727/0722-6691/5117}, \href
  {https://ui.adsabs.harvard.edu/abs/2019Msngr.175....3D} {175, 3}

\bibitem[\protect\citeauthoryear{{de Mink}, {Brott}, {Cantiello}, {Izzard},
  {Langer}  \& {Sana}}{{de Mink} et~al.}{2012}]{2012ASPC..465...65D}
{de Mink} S.~E.,  {Brott} I.,  {Cantiello} M.,  {Izzard} R.~G.,  {Langer} N.,
  {Sana} H.,  2012, in {Drissen} L.,  {Robert} C.,  {St-Louis} N.,   {Moffat}
  A.~F.~J.,  eds,  Astronomical Society of the Pacific Conference Series Vol.
  465, Proceedings of a Scientific Meeting in Honor of Anthony F. J. Moffat.
  p.~65

\bibitem[\protect\citeauthoryear{{de Wit}, {Testi}, {Palla}  \&
  {Zinnecker}}{{de Wit} et~al.}{2005}]{2005A&A...437..247D}
{de Wit} W.~J.,  {Testi} L.,  {Palla} F.,   {Zinnecker} H.,  2005, \mn@doi
  [\aap] {10.1051/0004-6361:20042489}, \href
  {https://ui.adsabs.harvard.edu/abs/2005A&A...437..247D} {437, 247}

\bibitem[\protect\citeauthoryear{{del Pino} et~al.,}{{del Pino}
  et~al.}{2022}]{2022ApJ...933...76D}
{del Pino} A.,  et~al., 2022, \mn@doi [\apj] {10.3847/1538-4357/ac70cf}, \href
  {https://ui.adsabs.harvard.edu/abs/2022ApJ...933...76D} {933, 76}

\bibitem[\protect\citeauthoryear{{van Rensbergen}, {Vanbeveren}  \& {De
  Loore}}{{van Rensbergen} et~al.}{1996}]{1996A&A...305..825V}
{van Rensbergen} W.,  {Vanbeveren} D.,   {De Loore} C.,  1996, \aap, \href
  {https://ui.adsabs.harvard.edu/abs/1996A&A...305..825V} {305, 825}

\makeatother
\end{thebibliography}




\appendix

\section{Alternative natal kicks}\label{sec:apx1}
\begin{figure*}
	\includegraphics[width=0.99\textwidth]{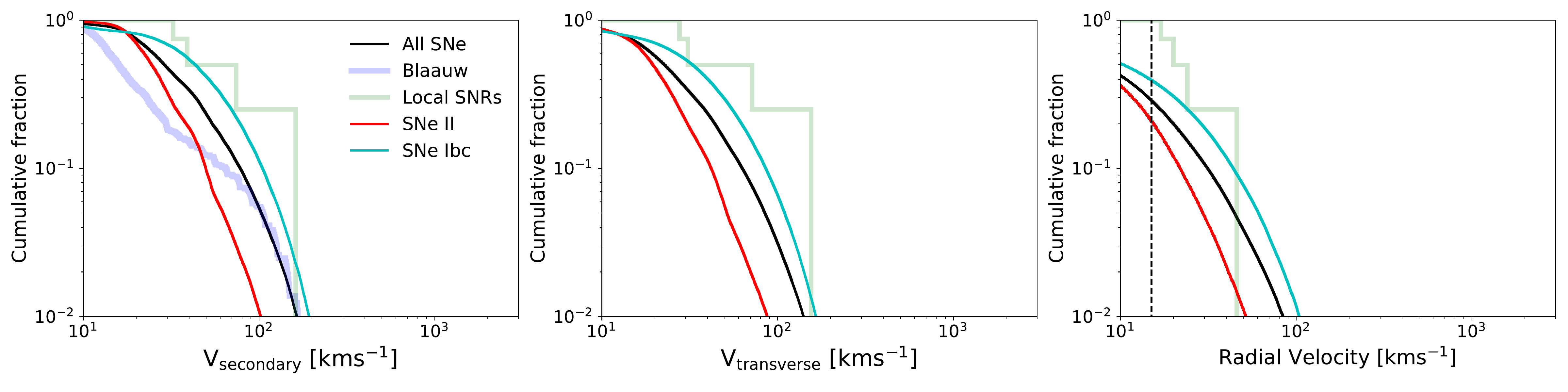}
   \caption{As in Figure \ref{fig:unbounds}, but with neutron star natal kicks from the distribution of \citet{2017A&A...608A..57V}, rather than \citet{2005MNRAS.360..974H}.}
    \label{apx:unbounds}
\end{figure*}

\begin{figure*}
	\includegraphics[width=0.99\textwidth]{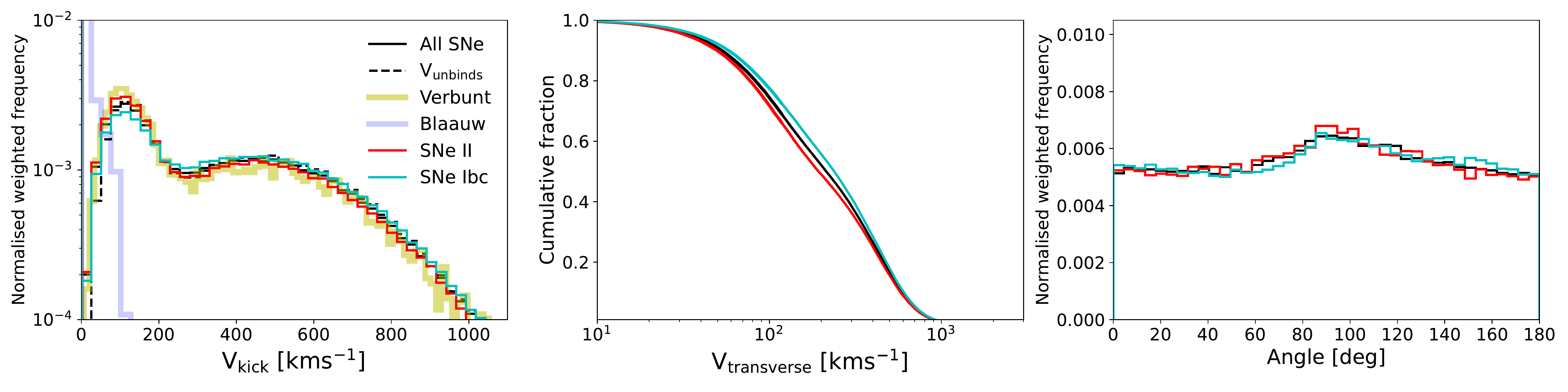}
   \caption{As in Figure \ref{fig:remnants}, but with neutron star natal kicks from the distribution of \citet{2017A&A...608A..57V}, rather than \citet{2005MNRAS.360..974H}.}
    \label{apx:remnants}
\end{figure*}

\begin{figure*}
	\includegraphics[width=0.99\textwidth]{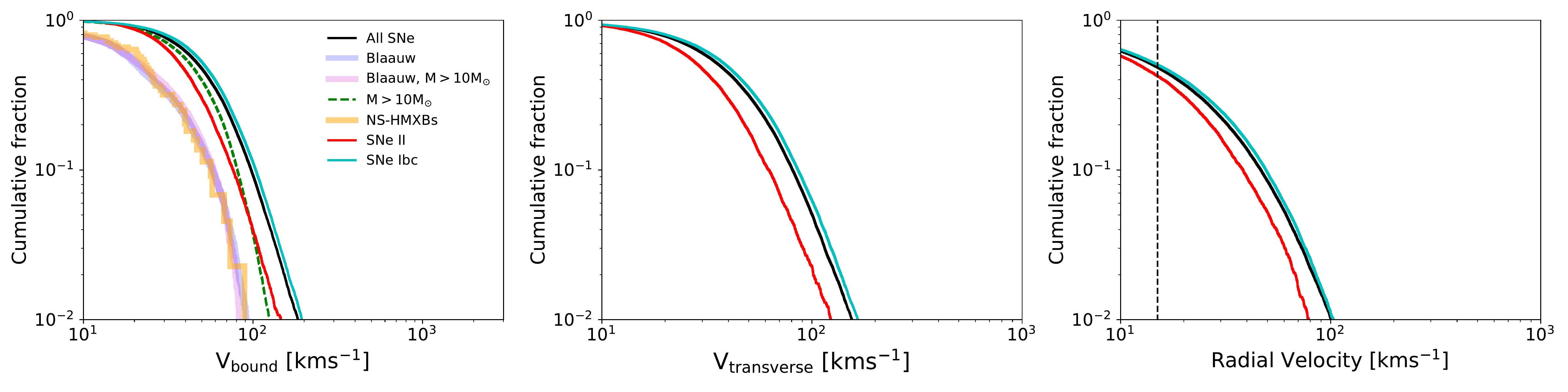}
   \caption{As in Figure \ref{fig:bounds}, but with neutron star natal kicks from the distribution of \citet{2017A&A...608A..57V}, rather than \citet{2005MNRAS.360..974H}.}
    \label{apx:bounds}
\end{figure*}

\begin{figure*}
	\includegraphics[width=0.99\textwidth]{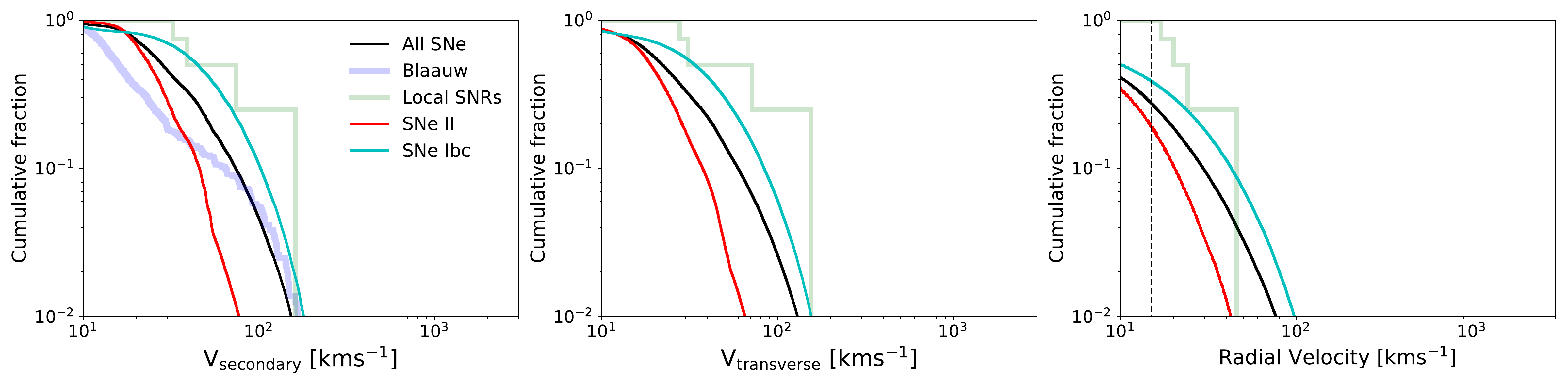}
   \caption{As in Figures \ref{fig:unbounds} and \ref{apx:unbounds}, but using the kick model of \citet{2016MNRAS.461.3747B}, with the model parameters of \citet{2022arXiv220802407R}.}
    \label{apx:brayunbounds}
\end{figure*}

\begin{figure*}
	\includegraphics[width=0.99\textwidth]{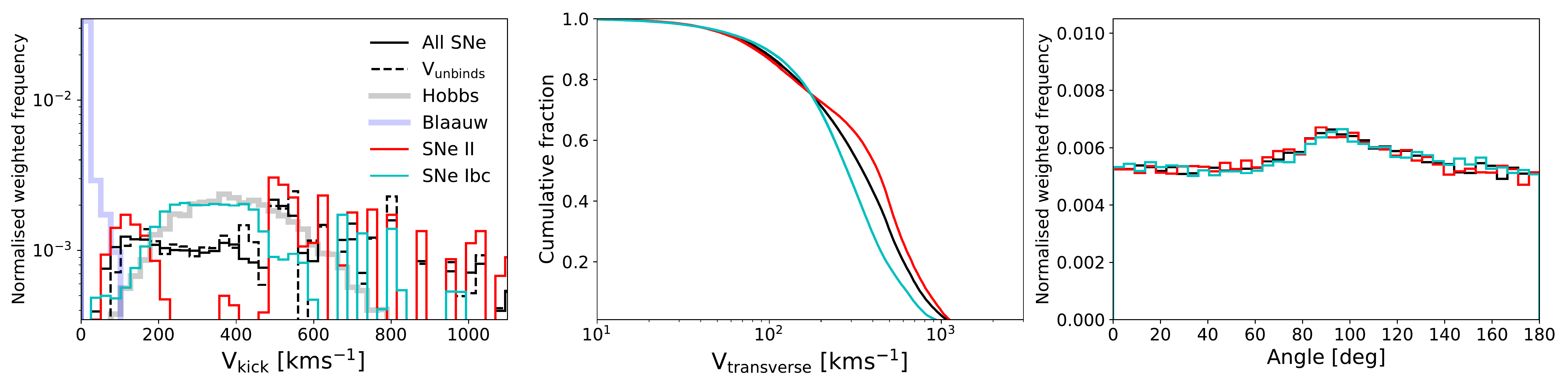}
   \caption{As in Figures \ref{fig:remnants} and \ref{apx:remnants}, but using the kick model of \citet{2016MNRAS.461.3747B}, with the model parameters of \citet{2022arXiv220802407R}. The \citet{2005MNRAS.360..974H} distribution is also shown on the left panel for reference.}
    \label{apx:brayremnants}
\end{figure*}

\begin{figure*}
	\includegraphics[width=0.99\textwidth]{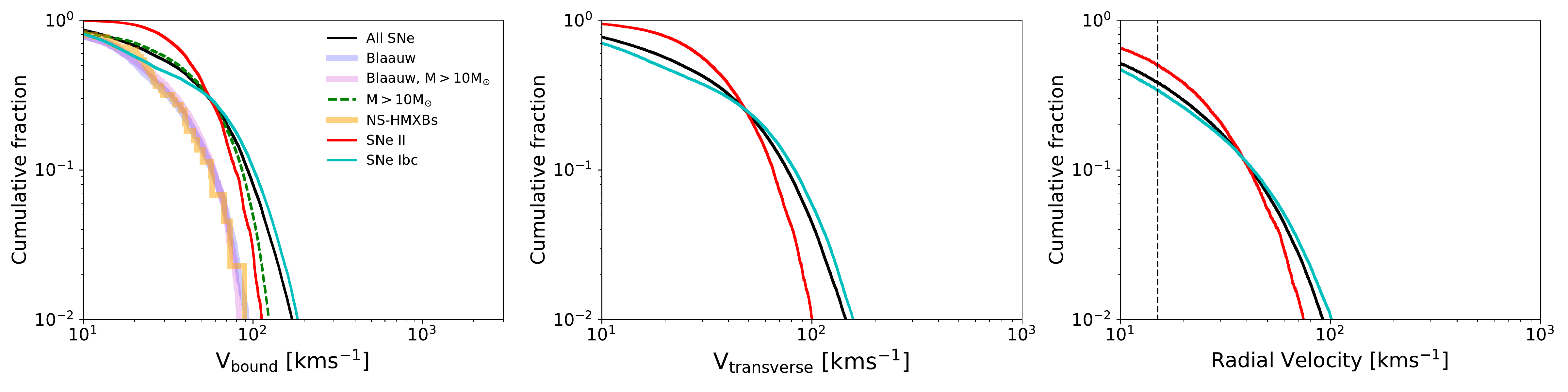}
   \caption{As in Figures \ref{fig:bounds} and \ref{apx:bounds}, but using the kick model of \citet{2016MNRAS.461.3747B}, with the model parameters of \citet{2022arXiv220802407R}.}
    \label{apx:braybounds}
\end{figure*}

For our fiducial models, we have assumed that neutron stars receive natal kicks following the \citet{2005MNRAS.360..974H} distribution. An alternative distribution, derived from young pulsars, is the bimodal distribution of \citet{2017A&A...608A..57V}. Alternative versions of Figures \ref{fig:unbounds}, \ref{fig:remnants} and \ref{fig:bounds} are provided in Figures \ref{apx:unbounds}, \ref{apx:remnants} and \ref{apx:bounds}, where \citet{2017A&A...608A..57V} kicks are assumed instead. Although it has been hypothesised that the low-velocity peak in this distribution corresponds to contributions from electron capture supernovae (ECSNe), given the uncertainty in which stars will produce ECSNe, we agnostically draw from the \citet{2017A&A...608A..57V}, as for \citet{2005MNRAS.360..974H}. In Figures \ref{apx:brayunbounds}, \ref{apx:brayremnants} and \ref{apx:braybounds}, we show equivalent results assuming the momentum-conservation kick model of \citet{2016MNRAS.461.3747B} and \citet{2018MNRAS.480.5657B}.

\section{Calculation of absolute magnitudes in JWST/NIRCam and Gaia filters}\label{sec:apx2}
\begin{figure}
	\includegraphics[width=0.99\columnwidth]{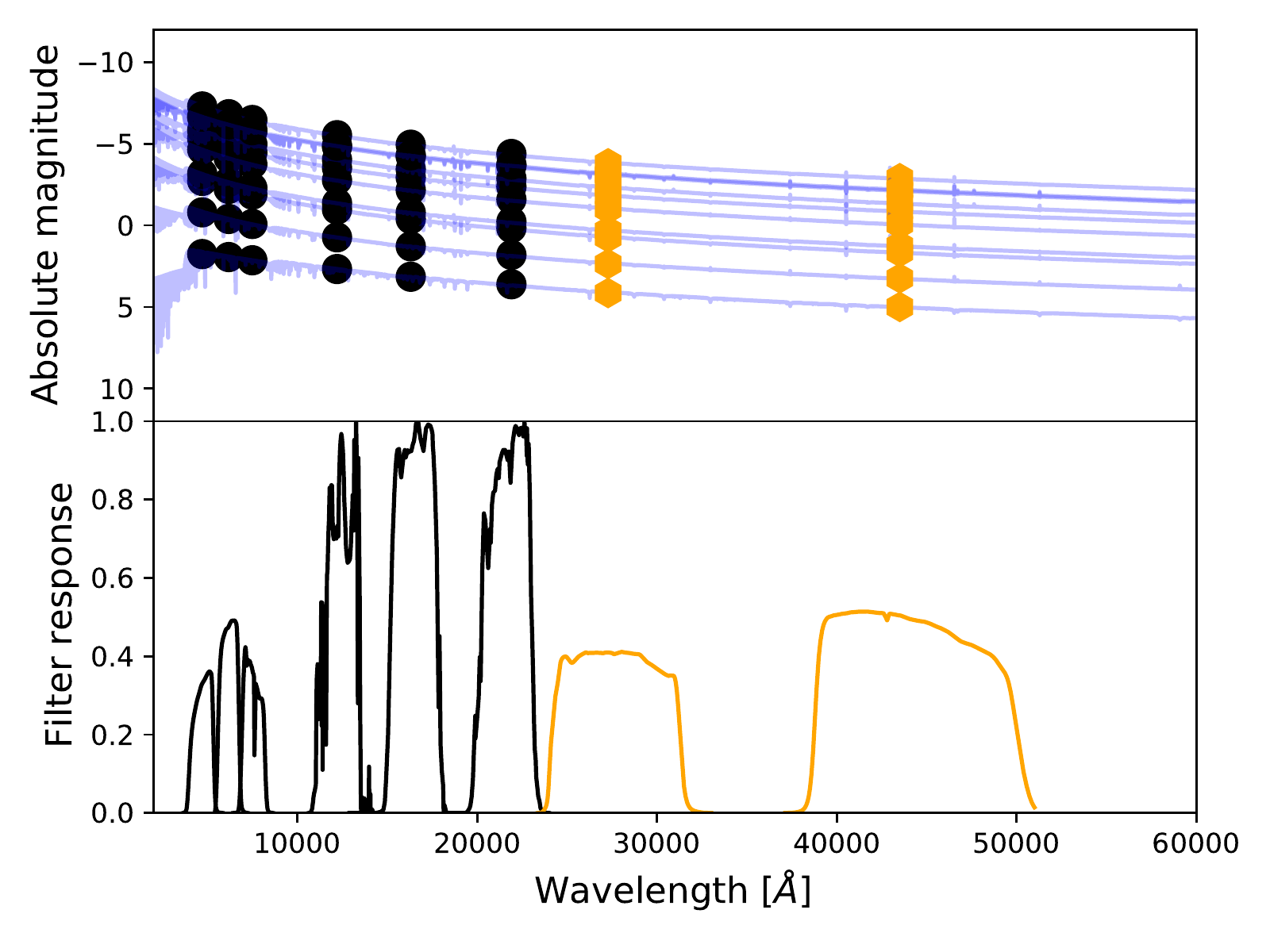}
    \caption{An example of our \jwst/NIRCam absolute magnitude calculations. Shown are ten randomly selected PHOENIX spectral fits to the synthetic photometry of unbound companions. Fitting is performed to BPASS $g$, $r$, $i$, $J$, $H$ and $K$ magnitudes (black points). The corresponding \jwst\ F277W and F444W magnitudes are then calculated from the best-fit PHOENIX model in each case (orange hexagons). The corresponding response curves used are shown in the lower panel \citep{2020sea..confE.182R}.}
    \label{apx:JWST}
\end{figure}

{\it JWST}: We first fit spectra to the BPASS synthetic photometry of the secondary star in the final timestep of the primary. To do this we use the Phoenix spectral library, whose spectra extend well beyond the 2-5$\mu$m range of interest for NIRCam. We fit for every model temperature and every log($g$). Since we are fitting models to models, we cannot perform a $\chi^2$ minimisation or similar, since the uncertainties are not quantified. Instead, we simply minimise the sum of the differences between each of the filters to determine the best fit. For converting filter magnitudes to fluxes across a wavelength range and vice versa, we use filter profiles from the SVO profile service \citep{2012ivoa.rept.1015R,2020sea..confE.182R}. For each BPASS secondary model we have a best-fit Phoenix spectrum, from which we then derive the \jwst\ F277W and F444W magnitudes. Figure \ref{apx:JWST} shows a selection of representative example fits, drawn at random from the unbound population.

\gaia: Since {\em Gaia} magnitudes are not a default output of BPASS, we instead approximate {\em Gaia} magnitudes ($G$, $G_{\rm BP}$ and $G_{\rm RP}$) from Johnson-Cousins V and I magnitudes using the fits of \citet{2021A&A...649A...3R}, with
\begin{equation}
    \begin{aligned}
    G = V - 0.01597 - 0.02809(V-I) - 0.2483(V-I)^{2} \\
    + 0.03656(V-I)^{3} - 0.002939(V-I)^{4}
    \end{aligned}
\end{equation}
for the $G$-band,
\begin{equation}
    \begin{aligned}
    G_{\rm BP} = V - 0.0143 + 0.3564(V-I) - 0.1332(V-I)^{2} \\
    +  0.01212(V-I)^{3}
    \end{aligned}
\end{equation}
for G$_{\rm BP}$ and
\begin{equation}
    \begin{aligned}
    G_{\rm RP} = V + 0.01868 - 0.9028(V-I) - 0.005321(V-I)^{2} \\
    - 0.004186(V-I)^{3}
    \end{aligned}
\end{equation}
for G$_{\rm RP}$.


\bsp	
\label{lastpage}
\end{document}